\documentclass{aa}  
\usepackage[colorlinks=true,allcolors=magenta]{hyperref}

\usepackage{graphicx}
\usepackage{txfonts}
\newcommand{\mnh}{N_\mathrm{H}}
\newcommand{\nh}{$\rm N_\mathrm{H}$}
\newcommand{\lognh}{$\rm \log\,N_\mathrm{H}$}

\begin{document} 
   \title{Identification of high-redshift X-ray active galactic nuclei in the 4XMM-DR11 serendipitous catalogue using DES data}

   \subtitle{Comparative analysis with optically-selected QSOs}
    \titlerunning{High-z X-ray AGN in the 4XMM-DR11 serendipitous catalogue }
    \authorrunning{E. Pouliasis et al.}

    \author{E.~Pouliasis\inst{1}\thanks{E-mail: epouliasis@noa.gr} 
    \and A.~Ruiz\inst{1}
    \and I.~Georgantopoulos\inst{1}
    \and A.~Akylas\inst{1}
    \and N.~A.~Webb\inst{2}
    \and F.~J.~Carrera\inst{3}
    \and S.~Mateos\inst{3}
    \and A.~Nebot\inst{4}
    \and M.~G.~Watson\inst{5}
    \and F.~X.~Pineau\inst{4}
    \and C.~Motch\inst{4}
}

    \institute{IAASARS, National Observatory of Athens, Ioannou Metaxa and Vasileos Pavlou GR-15236, Athens, Greece
    \and IRAP, Université de Toulouse, CNRS, CNES, 9 avenue du Colonel Roche, 21028 Toulouse, France
    \and Instituto de Física de Cantabria (CSIC-Universidad de Cantabria), Avenida de los Castros, 39005 Santander, Spain
    \and Université de Strasbourg, CNRS, Observatoire astronomique de Strasbourg, UMR 7550, 67000 Strasbourg, France
    \and School of Physics \& Astronomy, University of Leicester, University Road, Leicester LE1 7RH, UK}

   \date{Received  ; accepted }

% \abstract{}{}{}{}{} 
% 5 {} token are mandatory
 
  \abstract{X-rays provide a robust method in identifying active galactic nuclei (AGN). However, in the high-redshift Universe ($z \geq 3.0$), their space density is relatively low, and, in combination with the small areas covered by X-ray surveys, the selected AGN are poorly sampled. Deep optical/infrared data are essential for locating counterparts and determining redshifts. In this work, we leverage the \textit{XMM-Newton} 4XMM-DR11 serendipitous catalogue (1,240 $\rm deg^2$) alongside the extensive optical Dark Energy Survey (DES, 5,000 $\rm deg^2$) and the near-infrared VISTA Hemisphere Survey (VHS) to select one of the largest high-redshift X-ray AGN samples to date. Our analysis focuses on the overlapping area of these surveys, covering about 185 $\rm deg^2$. In addition, we aspire to compare the properties of the X-ray AGN with the optically-selected QSOs. For sources without spectroscopic data ($\rm \sim$80\%), we estimated the photometric redshifts using both SED fitting and machine-learning algorithms. Among the $\sim$65,000 X-ray sources in the 4XMM-DES-VHS area, we ended up with 833 $z \geq 3.5$ AGN (11\% having spec-z information) with high reliability and a fraction of outliers $\rm \eta \leq 10\% $. The sample completeness is $\sim$90\%, driven by the depth of DES data. Only $\sim$ 10\% of the X-ray selected AGN are also optical QSOs and vice versa. Our findings indicate an observed absorbed fraction ($\rm \log N_H~[cm^{-2}] \geq 23$) of 20–40\% for the X-ray AGN, significantly higher than that of optical QSOs. X-ray AGN exhibit fainter observed optical magnitudes and brighter mid-IR magnitudes than optical QSOs. Their median rest-frame SED shapes differ notably with optical QSOs being dominated by AGN emission in the UV-optical wavelengths. While the median SEDs of X-ray AGN suggest extinction in the UV-optical range, individual sources exhibit a wide range of spectral shapes, indicating significant diversity within the population. This analysis supports that X-ray- and optically-selected AGN represent distinct and complementary populations.}

   \maketitle
%
%-------------------------------------------------------------------

\section{Introduction}

Almost all massive galaxies in the Universe host a supermassive black hole (SMBH) in their center \citep{Magorrian1998,Kormendy2004,Filippenko2003,Barth2004,Greene2004,Greene2007,Dong2007,Greene2008}. When matter from the galaxies starts to accrete onto the SMBH, an enormous amount of energy is released across the electromagnetic spectrum (from radio emission up to X- and $\gamma$-rays). This constitutes the characteristic signature of Active Galactic Nuclei (AGN).

Despite the difference in the physical scale between the SMBH and the galaxy spheroid that is about nine times orders of magnitude, there is a tight correlation between the masses of SMBH and the galaxy bulge \citep{Silk1998,Magorrian1998,Ferrarese2000,Gebhardt2000}. In addition, many studies suggest a close interaction between the growth and evolution of galaxies and that of SMBHs \citep[][]{Silk1998,Granato2004,DiMatteo2005,Hopkins2006,Hopkins2008,Menci2008}. In particular, there is cumulative evidence for a similar cosmic evolution of the SMBH accretion and the star-formation (SF) rate density, both peaking at cosmic noon \citep[redshift z=1-3,][]{Madau2014,Aird2015,Harikane2022}. However, at higher redshifts ($z=4-5$), the AGN growth (in terms of black hole accretion density; i.e., total black hole growth rate per comoving volume) presents a much stronger decline over redshift compared to the SF rate density (SFRD; i.e., the amount of stars formed per year in a unit of cosmological volume). Thus, galaxy growth may precede the build up of their central SMBHs in the early Universe \citep{Aird2015}. Alternatively, the SMBH may have formed massive enough and thus they do not need high accretion rates to reach the local $\rm M_{BH}-M_\star$ relation. Additionally, it is possible that a population of primarily obscured AGN at high redshift remains undetected, which could contribute to the observed decline in AGN growth. Some simulations \citep{Ni2020,Habouzit2021,Ni2022,Gilli2022} indeed suggest the existence of such a population, which could help reconcile the observed trends in SMBH accretion and SFRD at high redshifts.

Most theoretical galaxy evolutionary models nowadays predict a regulating mechanism between the AGN power and the SF of the host galaxy leading to the formulation of the AGN-galaxy co-evolution paradigm \citep[e.g.][]{Hopkins2007,Lapi2014,Lapi2016}. In this scenario, merging systems are responsible for intense SF activity. The large fraction of gas reservoirs, then, is fuelled towards the SMBH and it initiates the AGN activity. This phase is characterized by growth of both the stellar mass of the galaxy and the SMBH. Since there are large quantities of gas, the AGN emission is absorbed and the sources appear as obscured AGN that last a long period of time. There are several studies indicating that a larger fraction of obscured AGN are found in mergers or post-merger galaxies \citep{Georgakakis2008,Satyapal2014,Lanzuisi2015,Ellison2019, Secrest2020}. X-ray observations of optically dual-AGN in late-state mergers also showed this trend \citep[e.g.][]{DeRosa2023}. More recent studies with the James Webb Space Telescope (JWST) support further the merging scenario. For example, \citet{Bonaventura2024} studied the morphology of the host galaxies of AGN with different obscuration levels and concluded that from the unobscured, moderately obscured to highly obscured AGN samples the fraction of disturbed galaxies increases from ~63\% to 95\%. As the SF and AGN consume, heat-up and expel gas, the AGN becomes less obscured, and, eventually, it becomes powerful enough to push away the surrounding material and turn into an unobscured type 1 AGN \citep{Ciotti1997,Hopkins2006,HopkinsHernquist2006,Somerville2008,Blecha2018}.

However, these theoretical models are in contrast to the standard unification model where the AGN obscuration depends only on the inclination angle related to the line of sight \citep[e.g.][]{Antonucci1993,Netzer2015,RamosAlmeida2017}. In addition, there are several studies that suggest that the obscured phase of the AGN is not the beginning in this scenario; the dust-obscured AGN are not the young quasars. For example, \citet{Georgantopoulos2023} compared the relative ages of the host galaxies of the absorbed and unabsorbed AGN and found that the absorbed AGN lie in intermediate-age galaxies between the young and old populations.

Hence, the physical processes governing the AGN-galaxy relationships are not yet fully understood and in order to validate the several evolutionary models of BHs and whether AGN affect their host galaxy properties, it is crucial to have a complete (as much as possible) census of the AGN demographics. In the last decades, even though a huge progress has been made in the construction of the BH demographics, especially at $z \leq 3$, there is a large fraction ($\sim$75\%) of highly obscured AGN predicted by the X-ray background (XRB) synthesis models that are currently undetected in X-rays \citep{Barchiesi2021}. The identification and classification of such sources becomes even more complicated in the high-z regime, since the obscuration (both the obscuring medium within the AGN, the so-called torus, and the larger-scale dust of the galaxy) increases with redshift reaching a plateau at $\rm z\sim 3-4$ \citep{Gilli2022,Signorini2023,Pouliasis2024}.

At redshifts higher than $z=3$, the expected density of luminous QSOs ($\rm M_{1450\AA, rest-frame} \leq -24$) is very low \citep[$\rm \sim1~Gpc^{-3}$,][]{DeRosa2014} and therefore large areas need to be probed. So far X-ray surveys have harvested limited high-z AGN samples because of the limited sky area covered. In particular, using both Chandra and XMM-Newton observations, several studies \citep[e.g.][]{Vito2014,Georgakakis2015,Marchesi2016highz,Vito2018,Khorunzhev2019,Pouliasis2022a, Peca2023} have identified a few hundred $z>3$ X-ray sources. In addition, the ongoing all-sky extended ROentgen Survey with an Imaging Telescope Array \citep[eROSITA,][]{Predehl2021} is expected to facilitate the search for high redshift AGN owing to its large grasp (field-of-view multiplied by effective area). After four years of operation, a few examples of very luminous $z>6$ AGN have been identified with the eROSITA detector with the majority of them also being radio-loud \citep[e.g.][]{Medvedev2020,Wolf2021,Wolf2023,Wolf2024}. The X-ray telescope onboard the {\it Gehrels/SWIFT} mission has provided yet another $z>6$ AGN \citep{BarlowHall2023}. \citet{Pouliasis2024} combining more than 600 Chandra and XMM-Newton high-z AGN, derived the most up to date X-ray luminosity function (XLF) indicating that about 60\% of the sources are obscured with hydrogen column densities of $\rm N_H \geq 10^{23}~cm^{-2}$. 

%%%%%%%%%%%%%%%%%%%%%%%%%%%%%%%%%%%%%%%%%%%%%%%%
Although their derived AGN space density and black hole accretion rate density (BHAD) are in agreement with the large-scale cosmological hydrodynamical simulations \citep[e.g.][]{Volonteri2016, Habouzit2022}, when considering the findings that X-ray AGN are hosted by massive galaxies, these results contrast with those obtained from mid-IR JWST data. For example, \citet{Yang2023}, using JWST observations, identified AGN through spectral energy distribution (SED) decomposition with a significantly higher density. In particular, the BHAD of the JWST-selected AGN is approximately 3 times larger than that of the X-ray-selected AGN. This discrepancy suggests the existence of an X-ray AGN population that may be missing due to either heavy obscuration or low intrinsic luminosity. However, more recently a similar JWST-MIRI study by \citet{Lyu2024}, which utilized deeper ancillary data and more accurate redshifts ($\sim$40\% spectroscopic) over a larger field, reported a significantly lower AGN number density of $\sim 6.3 ~arcmin^{-2}$, compared to the $\sim 14 ~arcmin^{-2}$ found by \citet{Yang2023}. \citet{Lyu2024} suggested that the latter higher AGN density may be due to contamination by low-mass galaxies and methodological differences, such as the treatment of PAH emission.

%%%%%%%%%%%%%%%%%%%%%%%%%%%%%%%%%%%%%%%

Furthermore, the ability of the optical/near-IR surveys to scan large areas in the sky (i.e., $\sim10^4\deg^2$) led to the discovery of a large sample of high-z broad-line QSOs increasing the diversity between the AGN populations in the early Universe. Shallow surveys such as SDSS \citep{Jiang2016}, the UKIRT Infrared Deep Sky Survey \citep[UKIDSS,][]{Mortlock2011}, the Canada-France High-redshift Quasar Survey \citep[CFHQS,][]{Willott2010} and the Panoramic Survey Telescope \& Rapid Response System \citep{Banados2018} played a significant role to select more than 400 confirmed spectroscopically high-z ( $z \gtrsim 5.5$) QSOs \citep{Fan2023}. However, only ($\sim$10\%) have been observed with deep X-ray observations, and out of those about 70\% have been detected \citep{Vito2019,Lusso2023,Wolf2024}. Deeper optical surveys with the Subaru Hyper Supreme Cam, HSC, \citep{Miyazaki2017} such as the HSC Subaru Strategic Plan Survey \citep{Aihara2018,Aihara2019} allowed the detection of AGN at redshifts $z=3-6$ at much fainter ($\rm >3~mag$) absolute magnitudes \citep{Akiyama2018,Niida2020,Matsuoka2022} compared to previous SDSS surveys. More recently, \citet{YangShen2023} released a sample of optically selected QSO candidates using the DES catalogue with high reliability ($\sim$95\%). Among them, there are more than 40,000 optically-selected QSO candidates at $z \geq 3$ in an area of 5,000 $\rm deg^2$.

\begin{figure*}
\begin{center}
\includegraphics[width=0.95\textwidth]{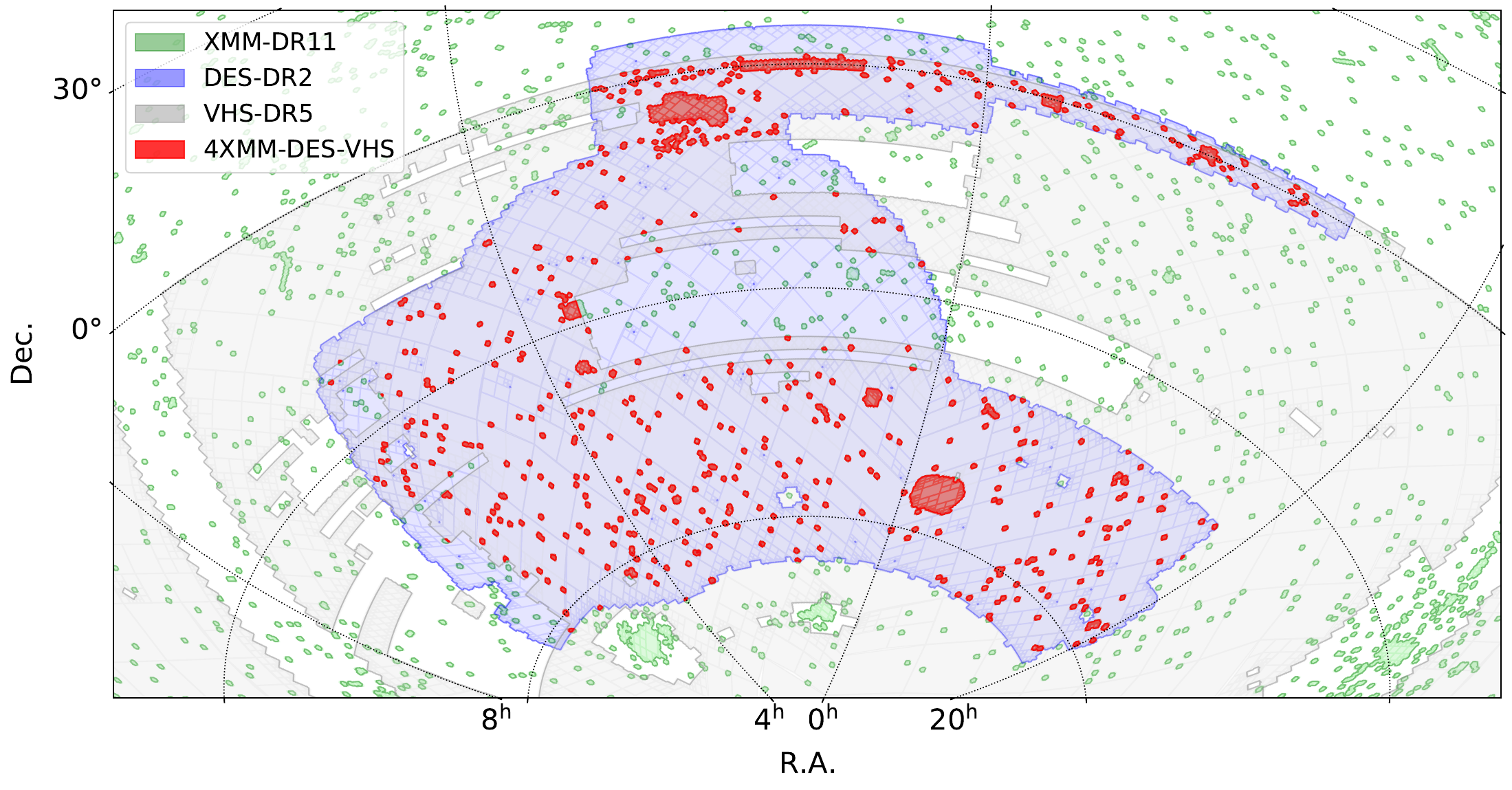} 
\end{center}
\caption{Survey footprint of the 4XMM-DES-VHS catalogue (red). In addition, we show separately the coverage of the 4XMM-DR11, DES-DR2 and VHS-DR5 surveys in the sky area of our interest.}\label{area}
\end{figure*}

Beside the limited sky areas covered by the X-rays, another obstacle is the lack of deep optical and/or near-IR photometry suitable for identifying the correct optical/NIR counterpart of the X-ray sources and to derive their photometric redshifts through SED fitting techniques. In this work, we aim at selecting high-z X-ray AGN using the wide-area 4XMM-DR11 catalogue \citep{Webb2020} and compare this sample to the optically-selected QSOs by \citet{YangShen2023}. In particular, we aim to investigate whether these two AGN populations are intrinsically different. The 4XMM-DR11 catalogue covers approximately 1,240 $\rm deg^2$ of the sky, while the second data release of the Dark Energy Survey \citep[DES-DR2,][]{Abbott2021} spans around 5,000 $\rm deg^2$, reaching faint magnitudes down to $\rm mag_i \sim 25$. Additionally, the VISTA Hemisphere Survey \citep[VHS,][]{Mcmahon2013} has been included in this analysis. The combined overlapping area between the X-ray and optical surveys, along with VHS, covers roughly 185 $\rm deg^2$, providing an excellent opportunity to directly compare X-ray AGN to optically selected QSOs.

The data used in our analysis are presented in Sect.~\ref{data}. Sections~\ref{highz} and \ref{final} describe the procedure followed to construct a reliable high-z X-ray sample discussing its completeness. In Sect.~\ref{sec:opticalqso}, we discuss the X-ray properties of the high-z sample along with the multi-wavelength properties derived using SED fitting, and we compare the observational properties of the X-ray AGN with respect to the optically-selected broad-line QSO sample. In Sect.~\ref{summary}, we summarise our results. Throughout the paper, we assume a $\Lambda$CDM cosmology with $\rm H\textsubscript{0}=70$ km s\textsuperscript{-1} Mpc\textsuperscript{-1}, $\rm \Omega\textsubscript{M}=0.3$ and $\rm \Omega \textsubscript{$\Lambda$}=0.7$ \citep{Komatsu2009}.

%%%%%%%%%%%%%%%%%%%%%%%%%%%%%%%%%%%%%%%%%%%%%%%%%%%%%%%%%%%%%%%%%%%%%%%%%
%%%%%%%%%%%%%%%%%%%%%%%%%%%%%%%%%%%%%%%%%%%%%%%%%%%%%%%%%%%%%%%%%%%%%%%%%
\section{Data}\label{data}

\subsection{4XMM-DES-VHS catalogue}

\begin{figure}
\begin{center}
\includegraphics[width=0.45\textwidth]{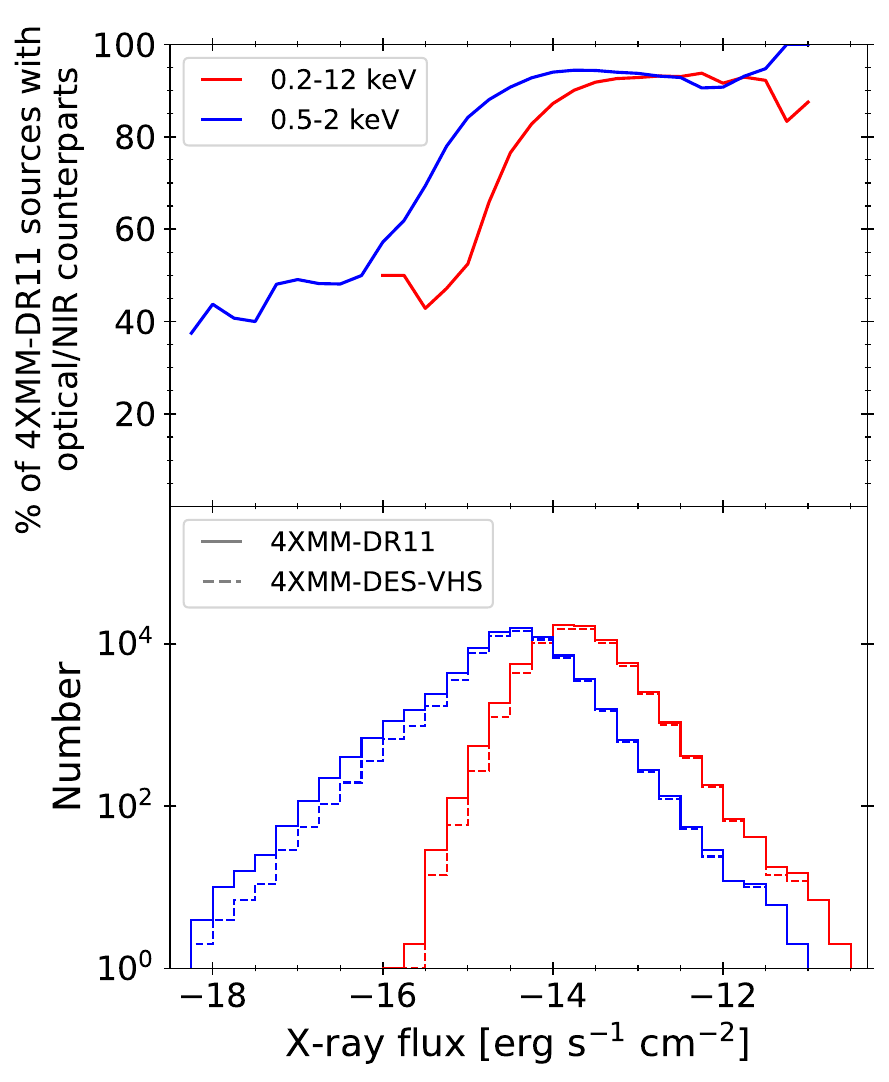} \\ 
\end{center}
\caption{Upper panel: Percentage of 4XMM-DR11 sources with an optical or near-IR counterpart as a function of X-ray flux in the soft (0.5-2 keV, blue) and the full band (0.2-12 keV, red). Lower panel: Flux distributions in the soft band and full bands of the 4XMM-DR11 and 4XMM-DES-VHS samples.}\label{x_compl} 
\end{figure}

\begin{figure}
\begin{center}
\includegraphics[width=0.5\textwidth]{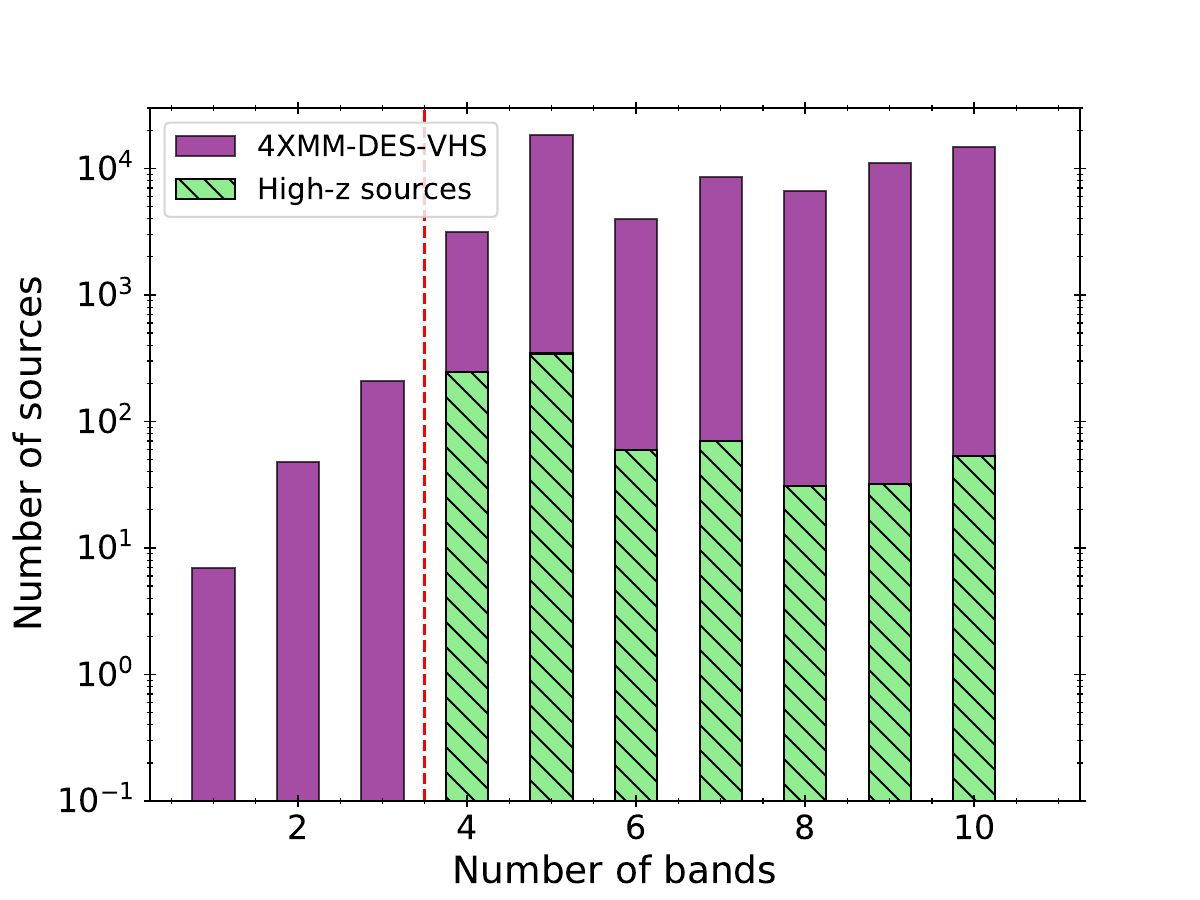} \\ 
\end{center}
\caption{Number of sources versus number of photometric bands for the 4XMM-DES-VHS (purple) and the final high-z (hatched, Sect.~\ref{final}) samples. The vertical dotted line corresponds to the cut at $\rm N \geq4$ we used in our analysis.}\label{Nbands} 
\end{figure}

In this work, we used the 4XMM-DR11\footnote{http://xmmssc.irap.omp.eu/Catalogue/4XMM-DR11/4XMM\_DR11.html} catalogue that contains all the serendipitous X-ray sources from the European Space Agency’s (ESA) \textit{XMM-Newton} observatory. 4XMM-DR11 includes data from 12,210 \textit{XMM-Newton} observations obtained by the European Photon Imaging Camera \citep[EPIC,][]{Struder2001,Turner2001}, that were made between 2000 February 3rd and 2020 December 17th and cover an energy range of 0.2-12 keV. The catalogue has been produced through the \textit{XMM-Newton} Survey Science Centre (SSC) on behalf of ESA. In total, there are 895,415 detections in these observations. The "slim" version of the catalogue contains 602,543 unique X-ray sources, where multiple detections of the same physical source appear only once. For the majority of the analysis in this paper, we utilized the "slim" version of the catalogue. However, we utilized the detection catalogue to calculate the total area curve and derive the cumulative number counts of the X-ray sources (Sect.~\ref{sec:lognlogs}), as well as to determine the X-ray spectral properties of the high-z sample (Sect.~\ref{sec:xrays}).

The classification of X-ray sources used in this work was performed by the XMM2ATHENA project\footnote{http://xmm-ssc.irap.omp.eu/xmm2athena/wp-content/uploads/2023/04/DeliverableD8.2.pdf}. Following the classification work using a newly developed Naive Bayes classifier, presented in \citet{Tranin2022}, the XMM2ATHENA team further refined the algorithm and applied it to the 4XMM catalogue \citep{Webb2020}. This updated algorithm is capable of identifying sources as AGN or galaxy, extragalactic and galactic X-ray binaries (XRBs), stars, cataclysmic variables (CVs), and extended sources, enabling the distinction between Galactic and extragalactic sources. To enhance reliability, the classification was augmented with data from multi-wavelength catalogues and validated through manual screening and testing with sources of known types. Among the sources with a secure classification (99.9\%), we selected only those flagged as AGN or QSO, representing nearly 95\% of the total X-ray sources. The false-positive rate of sources misclassified as AGN is estimated to be approximately 5\%. However, in the next section this point will be discussed further as in the high-z regime, this percentage increases. The sample of point-like X-ray selected AGN/QSOs inside the 4XMM-DES-VHS footprint consists of 75,782 sources.

For the optical photometry, we used the second data-release of the dark energy survey \citep[DES-DR2,][]{Abbott2021} that contains photometry in five broad bands (g, r, i, z and Y), while for the NIR photometry (J, H, Ks bands) we made use of the fifth release of the VISTA Hemisphere Survey \citep[VHS-DR5,][]{Mcmahon2013}. We only used the DES-DR2 bands with reliable photometry: $Flags \leq 4$ (well-behaved objects), $imaflags_{iso}=0$ (no missing/flagged pixels in the source in all single epoch images) and $N_{iter}=0$ that ensures no missing imaging data. In addition, the 4XMM-DES-VHS footprint has taken into account the DES DR2 masking regions \citep{Abbott2021} that exclude sources near bright stars that may be affected by diffraction spikes or high-background counts, near bleed trails, and artificial satellite streaks. The selection of the optical/NIR counterparts of the X-ray sources was achieved under the framework of the H2020 XMM2ATHENA project \citep{Webb2023} using the ARCHES statistical analysis \citep{Motch2017} for multi-wavelength cross-correlation and cross-identification that is based on astrometric criteria \citep{Pineau2017}. By selecting the counterparts with the highest probability being the correct one, we ended up with 66,798 X-ray sources with an optical and/or NIR counterpart. The completeness with respect to the whole X-ray catalogue inside the DES/VHS footprint is 88\%.

In Fig.~\ref{x_compl} (upper panel), we show the percentage of the 4XMM-DR11 sources with an optical/NIR counterpart versus the X-ray flux in the full band (0.2-12 keV, red). The lower panel presents the flux distributions of the 4XMM-DR11 and 4XMM-DES-VHS samples. It is important to note here that the 4XMM catalogue used in our analysis contains detections with a significance greater than $\sim3 \sigma$ in the full band, which corresponds to a maximum likelihood of 6 \citep{Webb2020}. For 73\% of the sources in our sample, the significance is greater than $5 \sigma$ (maximum likelihood of 14). For reference, we also present the percentage of the soft band (0.5-2 keV, blue) sources with  an optical/NIR counterpart that was used for the high-z sample incompleteness estimation (Sect.~\ref{completeness}). In the bright end ($\rm S_{0.5-2~keV} > 10^{-15} erg~s^{-1}~cm^{-2}$, which represent 85\% of 4XMM-DR11), the completeness is about 90\%, however at fainter fluxes it drops down to 50\%. Regarding the tail of sources with a soft X-ray flux below $1 \times 10^{-16} \, \mathrm{erg \, cm^{-2} \, s^{-1}}$, these constitute less than 2\% of the full sample, and for more than half of these, the significance remains above $5 \sigma$. As mentioned above, the 4XMM catalogue was build upon the full band detections, hence we expect some hard sources to have very faint fluxes in the soft band.

Furthermore, we complemented the multi-wavelength catalogue with mid-IR photometry from the Wide-field Infrared Survey Explorer mission \citep[WISE;][]{Wright2010}. In particular, we cross-matched the optical/NIR counterparts to the X-ray sources with the ALLWISE source catalogue \citep{Cutri2013} using a radius of 2" following \citet{Fotopoulou2016b}. More information in the datasets and the cross-matching procedure will be available in Ruiz et al. (in prep.). We acknowledge that no explicit correction for the different resolutions was applied during the positional cross-matching of photometric data points. Although this may introduce some uncertainties in identifying the correct MIR counterparts, particularly for blended sources in WISE images, the overall effect on the photometric redshift solutions is expected to be minimal. However, we estimated that only about 5.4\% of WISE sources are likely blended with multiple counterparts within the 2" matching radius, based on a self-crossmatch of the optical/NIR catalogue. This low blending fraction minimizes the potential impact on the photo-z solutions, as the majority of sources remain uncontaminated by resolution-related effects. Finally, in order to construct reliable SEDs with the available photometry, to derive the photometric redshifts and also to decompose the AGN and the galaxy components, we required for all sources to have at least four bands available. Figure~\ref{Nbands} presents the number of sources versus the number of bands available for the 4XMM-DES-VHS sample. The vertical line corresponds to the threshold on number of bands we set ($\rm N \geq4$). This cut excludes almost 5\% of the 4XMM-DES-VHS sample. The majority of the sources (94\%) do have detections in all five DES bands. This is in contrast with the infrared bands, where 60\% (52\%) of the sources have at least one NIR (mid-IR) band available. This mainly is due to the different depths of the optical compared to near- and mid-IR photometry. Hence, we ended up with a sample of 66,524 X-ray sources in the 4XMM-DES-VHS footprint to be used in the selection of high-z sources.

\subsection{Spectroscopic redshifts}\label{specz}

Inside the 4XMM-DES-VHS area, there is a plethora of optical spectroscopic information originating from different surveys targeting extra-galactic sources including both AGN and galaxies. For this work, we made use of the Millions of Optical-Radio/X-ray Associations (MORX) Catalogue \citep{Flesch2016,Flesch2024}. In particular, we used the second version of the MORX catalogue that includes 3,115,575 optical objects with radio or X-ray associations. About 20\% of the MORX sources have a spectroscopic redshift. The 4XMM-DR13 catalogue \citep{Webb2020} is one out of the several X-ray surveys used to build this catalogue. We cross-matched this dataset to the 4XMM-DES-VHS catalogue using the optical/NIR coordinates and a searching radius of 1". Among the 32,353 MORX associations, there are 13,424 X-ray sources with available spectroscopy.

\subsection{Photometric redshifts}

\begin{figure}
\begin{center}
\includegraphics[width=0.45\textwidth]{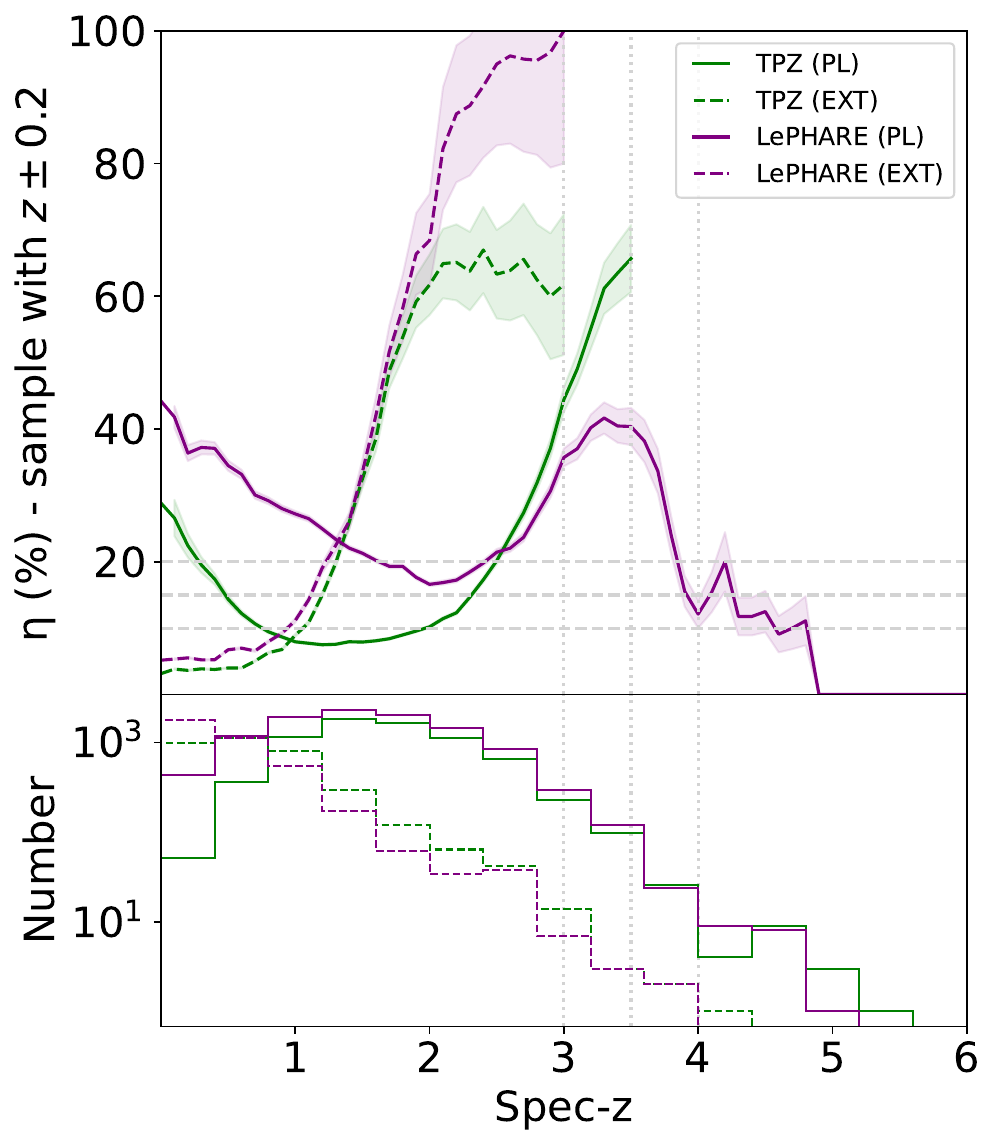}  
\end{center}
\caption{Percentage of outliers versus spectroscopic redshift (upper panel). The solid (dashed) lines correspond to point-like (extended) sources for the TPZ (green) and LePHARE (purple) results. Shaded regions represent the uncertainty in the percentage of outliers at each spec-z bin, calculated as the percentage of outliers divided by $\sqrt{N}$, where N is the number of sources in each bin. Larger sample sizes result in narrower shaded regions. The bottom panel shows the distribution of the training sample with spectroscopic redshifts in each case.} \label{plot_statistics_paperPLxonwTPZb.png}\label{photoz_accuracy}
\end{figure}

\begin{figure}
\begin{center}
\includegraphics[width=0.45\textwidth]{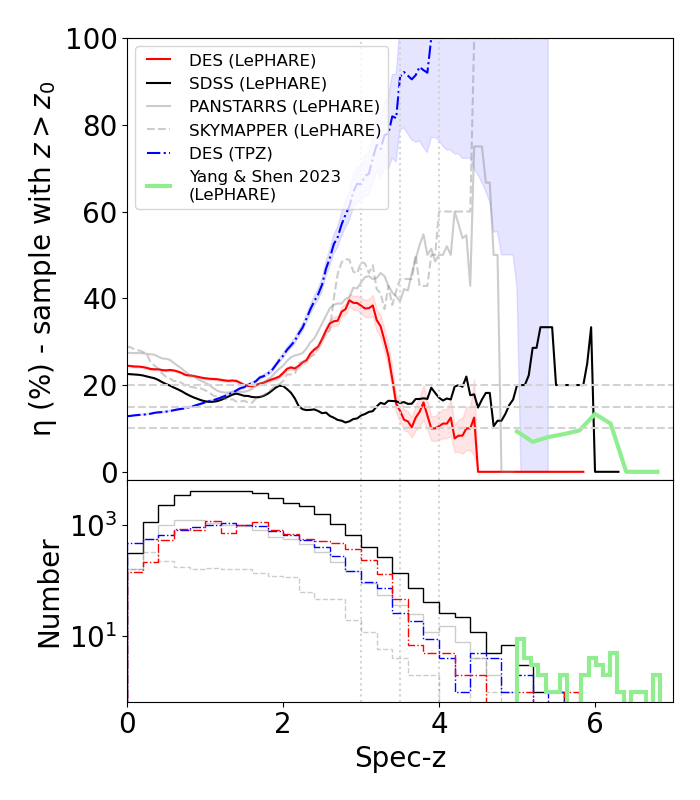}  
\end{center}
\caption{Percentage of outliers versus spectroscopic redshift (upper panel) for the point-like sources. At each redshift, we calculate the percentage of outliers for all sources above this redshift. The bottom panel show the distribution of the training samples with available spectroscopic information. The performance in the DES field using TPZ (blue, dashed-dotted) and LePHARE (red, solid) is compared across different redshifts. For reference, we show the performance of LePHARE algorithm for the 4XMM-DR13 sources using different optical photometry adopted from Ruiz et al. (in prep.), such as SDSS (black, solid), PANSTARRS (gray, solid) and SKYMAPPER (gray, dashed). Finally, we present the LePHARE performance of $z \geq 5$ sources of \citet{YangShen2023}.}\label{photoz_statistics} 
\end{figure}

In Ruiz et al. (in prep.), we derived photometric redshifts for the 4XMM-DES-VHS sources using both Machine-learning (ML) algorithms and SED fitting methods. Specifically, we applied the Trees for Photo-z (TPZ) algorithm \citep{Carrasco2013}, which is based on random forest techniques, and the LePHARE code \citep{Arnouts1999,Ilbert2006}. For LePHARE, it is essential to run the code with different SED templates depending on the morphological type of the source (point-like or extended) to prevent parameter degeneracies in the models \citep[e.g.][]{Pouliasis2024}. To classify the sources according to their morphological type, we used the parameter $\rm EXTENDED\_COADD$ provided in the DES-DR2 catalogue. Following \citet{Abbott2021}, we selected point-like and extended sources using $\rm EXTENDED\_COADD<2$ and $\rm EXTENDED\_COADD\geq2$, respectively. In addition, we applied the appropriate luminosity priors in each case. In particular, we used an absolute magnitude of $\rm -23 < M_{DES g} < -8$ for the extended and $\rm -30 < M_{DES g} < -20$ for the point-like sources \citep[][and references therein]{Pouliasis2024}. To assess the accuracy of photometric redshifts derived from TPZ and LePHARE, we used the sub-sample of X-ray sources with available spectroscopic redshifts (Sect.~\ref{specz}), from which we calculated the percentage of outliers, $\eta$ \citep{Ilbert2006,Laigle2016}, defined as:
\begin{equation}
\eta (\%)=\frac{N_{\rm outliers}}{N_{\rm total}}\times100,
\end{equation}
where $N_{\rm total}$ is the total number of sources and $N_{\rm outliers}$ is the number of the outliers. An object is defined as an outlier if it has $|\Delta z|/(1+z_{\rm spec})>0.15$, where $\Delta z=z_{\rm phot}-z_{\rm spec}$. Figure~\ref{photoz_accuracy} illustrates the percentage of outliers as a function of redshift for both point-like and extended sources using the TPZ and LePHARE algorithms. For each spectroscopic redshift, we calculate the percentage of outliers within the range $z_{\rm spec} \pm 0.2$. This range is used to evaluate the number of outliers within a window around the spectroscopic redshift. It is independent of the $|\Delta z|$ definition, which measures the difference between photometric and spectroscopic redshifts for individual sources.

ML algorithms have proven highly effective and accurate for estimating photometric redshifts using broad-band photometry, but they require large training sets of sources with spectroscopic redshifts. Consequently, ML algorithms like TPZ are very powerful at $z<3$, but their performance diminishes beyond this redshift due to a lack of spec-z data. To overcome this limitation, SED fitting algorithms like LePHARE can be used, which is why we decided to rely on LePHARE's photometric redshift estimates for all X-ray sources without spec-z information. However, in the next section we make use of the secure and accurate TPZ results to filter out low-z galaxies from the LePHARE high-z sample, ensuring a more reliable selection of high-redshift sources.

%%%%%%%%%%%%%%%%%%%%%%%%%%%%%%%%%%%%%%%%%%%%%%%%%%%%%%%%%%%%%%%%%%%%%%%%%
%%%%%%%%%%%%%%%%%%%%%%%%%%%%%%%%%%%%%%%%%%%%%%%%%%%%%%%%%%%%%%%%%%%%%%%%%

\section{Selection of high-z sample}\label{highz}

\subsection{Identification of high-z sources}\label{identificationhighz}

\begin{figure}
\begin{center}
\includegraphics[width=0.45\textwidth]{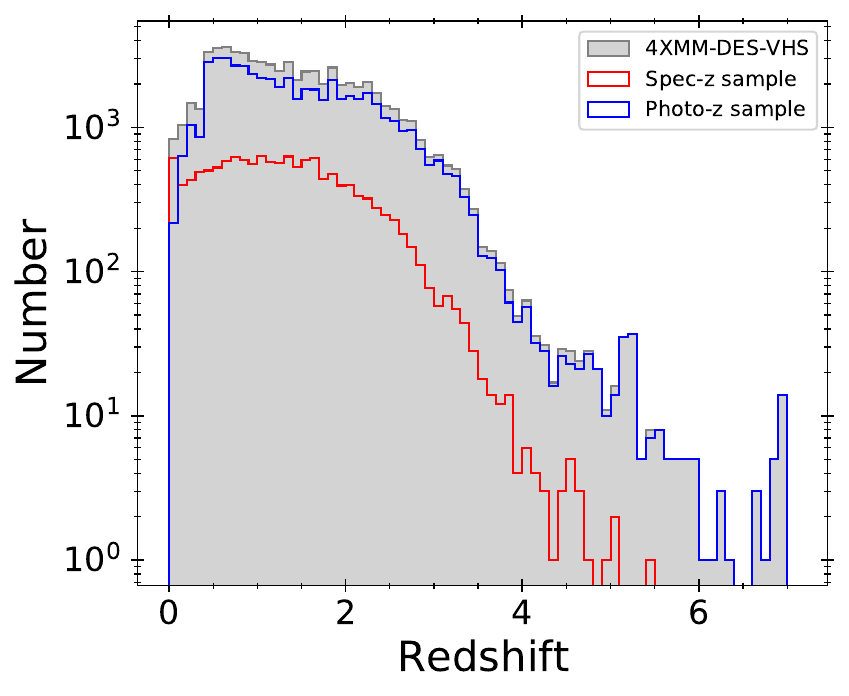} 
\end{center}
\caption{The redshift distribution of the 4XMM-DES-VHS sample used in our analysis (gray filled). We highlight the spectroscopic and photometric redshift samples as indicated in the legend.}\label{hist_z} 
\end{figure}

Using the MORX spectroscopic catalogue mentioned in Sect.~\ref{specz}, we select all the X-ray sources with $z \geq 3.5$. Out of 13,424 X-ray sources with spectroscopic redshift, there are 92 AGN at $z \geq 3.5$. Among them, 30 sources are found at $z \geq 4$, while three sources at $z \geq 5$. For sources that lack spectroscopic information, we use the photometric redshift estimations.

The threshold of $z=3.5$ in selecting high-redshift sources was defined in order to have a low percentage of outliers in the photometric-redshift sample. In Fig.~\ref{photoz_statistics}, we plot the percentage of outliers vs the spectroscopic redshift (upper panel, red line). At each spec-z value, we estimate the percentage of outliers for all the sources above this redshift. The bottom panel shows the histogram of the spec-z sources. Around $z=3$, we observe a noticeable bump corresponding to an increased percentage of outliers, which is expected due to template degeneracies. This issue is primarily caused by the absence of the u-band in the optical (DES) photometry. For comparison, we also show the results for 4XMM-DR13 sources with SDSS photometry (Ruiz et al., in prep.), which includes the u-band, and the bump disappears. Consequently, in the absence of the u-band, one should target sources beyond $z \sim 3.2$ to reliably select high-redshift objects. For sources with $z>3.5$, DES provides statistically higher accuracy compared to SDSS due to its greater depth (the median $5\sigma$ depths of DES-DR2 and SDSS in i-band are $\rm i_{DES}=24.5 ~mag$ and $\rm i_{SDSS}=22.2 ~mag$, respectively). To further emphasize this point, we also present results using PAN-STARRS \citep{Chambers2016} and Skymapper \citep{Wolf2018, Onken2019} optical photometry (Ruiz et al., in prep.). Both of these surveys, like DES, lack the u-band but are also much shallower. In addition to the bump observed around $z=3$, we see an increasing trend in the percentage of outliers for $z>3.5$, highlighting the importance of deep optical data.

For the DES dataset, the redshift accuracy improves significantly at $z>3.5$, with the outlier fraction dropping below 12\%. This percentage further decreases to approximately 10\% for sources with $z>4$ and approaches zero for sources with $z>5$. This behavior is expected, as at higher redshifts ($z>4$), low-redshift galaxies are less likely to contaminate the sample due to the spectrum shift. However, since our spectroscopic sample for $z>5$ is limited, we followed a similar approach to \citet{Saxena2024} to evaluate the performance of the LePHARE algorithm in this regime. In particular, they used the CIRCLEZ algorithm to derive photometric redshifts for X-ray selected AGN and relied on an optical QSO sample for training in the absence of high-z X-ray spectroscopic data. Similarly, we tested the performance of LePHARE at $z>5$ using the \citet{YangShen2023} sample. Specifically, we ran LePHARE on 43 sources with $z>5$, of which only four were identified as outliers. Figure 5 shows the fraction of outliers versus redshift for this sample, where the outlier fraction is approximately 10\%, dropping further at $z>6$. This represents an upper limit, as the sample consists of optically selected QSOs dominated by central engine emission. In contrast, as discussed later, we expect a lower outlier fraction for X-ray-selected sources, where host galaxy emission is more prominent.

At the end, out of the 66,524 X-ray sources in our sample, there are 972 sources with $z \geq 3.5$ obtained by the LePHARE algorithm without spectroscopic information.

\subsection{Low-z galaxy interlopers}\label{lowz}

From the percentage of outliers for the test sample discussed above, we would expect less than 12\% of the sources with photometric high redshifts to be low-z galaxy interlopers. As noted earlier, ML algorithms for photometric redshift estimation are highly accurate for $\rm z<3$ (Fig.~\ref{photoz_statistics}, left). To address this, we utilized the most reliable TPZ results, where reliability is defined by the peak strength (PS) value of the probability density function of redshift \citep{Ruiz2018}. Hence, we used all sources with PS > 0.7 (Ruiz et al., in prep.) to identify and exclude these interlopers from the high-redshift sample. This threshold was chosen, as investigated by \citet{Ruiz2018}, to balance the trade-off between sample purity and completeness. While this criterion does not entirely eliminate the possibility of secondary redshift solutions at lower redshift, it significantly reduces their likelihood, thereby enhancing the reliability of our high-redshift sample.

In total, we identified 161 X-ray sources, which accounts for about 17\% of the photometric high-z sample which serves as a lower limit. Figure~\ref{hist_z} shows the redshift distribution for the whole sample (shaded gray area). The red and blue histograms represent the spectroscopic and photometric redshift samples, respectively. In the high-redshift regime, there are two abnormal overdensities at $z \sim 5.2$ and $z \sim 6.9$. We examined all the information available and found that the first one is due to stellar objects misclassified as AGN. The second overdensity is mainly due to imaging issues and/or sources being near very bright objects that passed our criteria initially Sect.~\ref{data}. The next two sections describe the removal of these objects from our high-z sample.

\subsection{Stellar contamination}\label{stars}

\begin{figure}
\begin{center}
\begin{tabular}{c}
\includegraphics[width=0.48\textwidth]{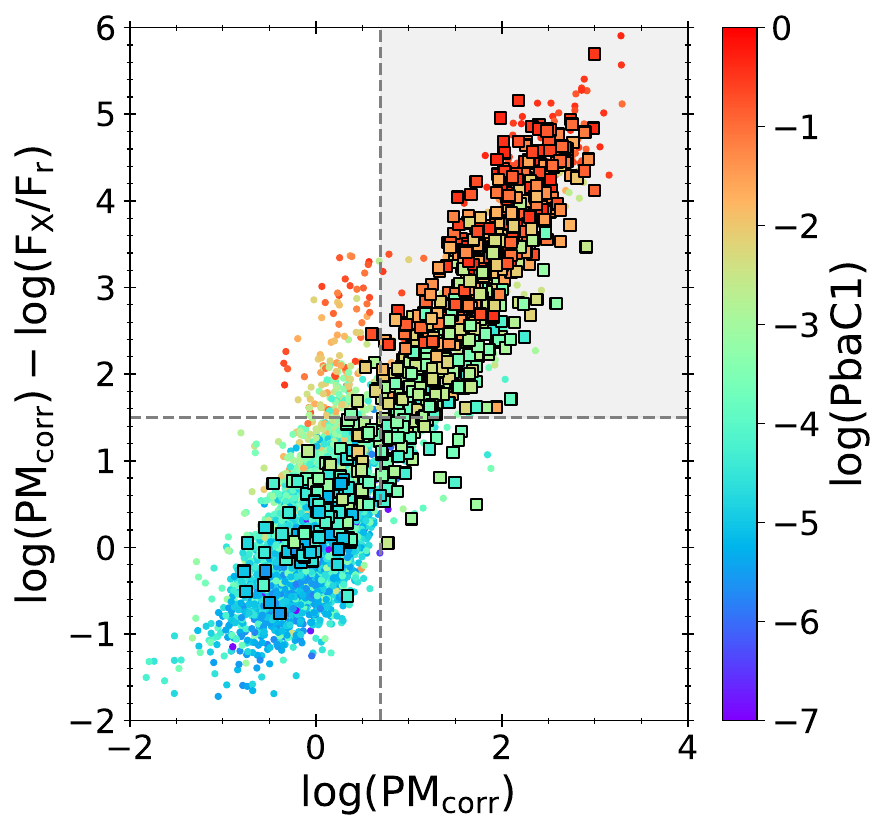} \\ 
\includegraphics[width=0.48\textwidth]{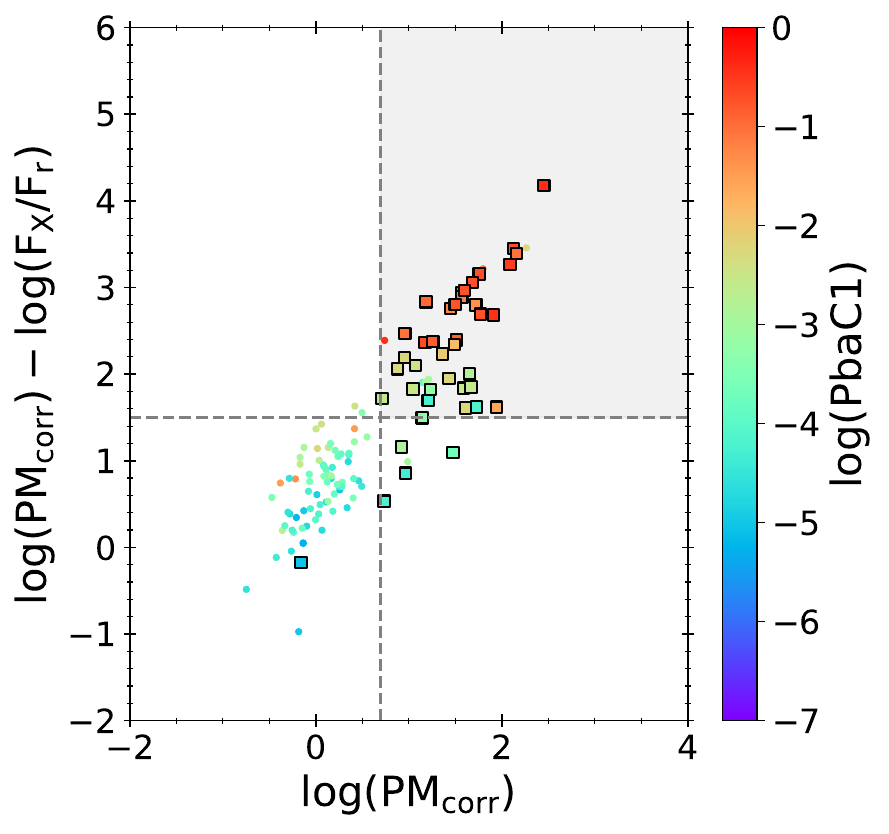}  
\end{tabular}
\end{center}
\caption{$\rm log(PM_{corr})-log(F_X/F_r)$ versus the corrected Gaia proper motion in logarithmic scale, $\rm log(PM_{corr})$, where $\rm F_X/F_r$ is the X-ray to optical flux ratio. The dashed lines correspond to the Gaia criteria to select stars. The sources are colour-coded with the probability of a source being a star from the classification algorithm \citep{Tranin2022}, the $\rm PbaC1$ class. The squares indicate sources with $\Delta \chi^2 \geq 5$, where $\rm \Delta \chi^2$ is the difference in the $\rm \chi^2$ between the best fit of AGN/QSO versus stellar templates. The upper and lower panels correspond to the total and the high-z samples, respectively.}\label{gaia} 
\end{figure}

Although we have initially selected the AGN/QSO class for our analysis, we expect about 5\% of these sources to be stars and misclassified as AGN/QSO \citep{Tranin2022}. In order to remove the remaining stellar population contaminating our high-z AGN sample (first overdensity at $z \sim 5.2$), we utilised multiple criteria for sources with point-like morphology in the optical images. Firstly, we used the Gaia source catalogue \citep{Gaia2016,Gaia2023} and removed all the sources with high proper motion ($\rm PM \geq 5$). In order to take into account the uncertainties ($\rm pmra\_{error}$ and $\rm pmdec\_{error}$) and the correlation ($\rm pmra\_pmdec\_{corr}$) of the PM components along the equatorial coordinates we followed the definition of \citet{Carnerero2023}:
\begin{equation}
    \rm PM_{corr} = \sqrt{\frac{\alpha^2 + \beta^2 - 2 \alpha \beta \gamma}{1-\gamma^2}}\geq 5,
\end{equation}
where $\rm \alpha = pmra/pmra\_{error}$, $\rm \beta = pmdec/pmdec\_{error}$, and $\rm \gamma = pmra\_pmdec\_{corr}$. In addition, we excluded sources with $\rm log(PM_{corr}) - log(F_X/F_r) \geq 1.5$, where $\rm F_X/F_r$ represents the X-ray (0.5-2 keV band) to optical flux ratio. This criterion is based on the $\rm log(PM_{corr})$ and $\rm log(F_X/F_r)$ plane, where stellar and AGN populations are distinct. The value of 1.5 corresponds to the optimal separation between these two populations.

However, the Gaia matches to our catalogue consist of less than 25\% of the whole 4XMM-DES-VHS sample. Hence, for sources without Gaia information, we made use of the probability of a source being a star from the XMM2ATHENA classification algorithm, $\rm PbaC1$ \citep{Tranin2022}, and further the difference in the $\rm \chi^2$ between the best fit of AGN/QSO versus stellar templates. For the latter, we ran the LePHARE algorithm with stellar templates and calculated the difference as $\rm \Delta \chi^2 = \chi^2_{AGN}-\chi^2_{star}$. A negative $\rm \Delta \chi^2$ indicates that the AGN model provides a better fit to the data compared to the stellar model, while a positive or small $\rm \Delta \chi^2$ suggests that the stellar model is preferred or both models fit equally well. The significance of the $\rm \Delta \chi^2$ is assessed by comparing it to a threshold, beyond which we can confidently reject the stellar model in favor of the AGN model. To define the thresholds for both these criteria, in Fig.~\ref{gaia} (top panel) we plot $\rm log(PM_{corr})-log(F_X/F_r)$ versus $\rm log(PM_{corr})$. The dashed lines correspond to the Gaia criteria mentioned above to select a reliable stellar sample. The stellar population is located in the top right corner of this plot (shaded region). The bottom panel of Fig.~\ref{gaia} correspond to the high-z sample. Based on these, we set a cut-off at $\rm \Delta \chi^2 = 5$ (black squares) and also a $\rm logPbaC1=-1.25$. Sources that have values greater than these are considered to be stars. By using all four criteria, we were able to exclude in total 1,854 stars, that is about 2\% of the point-like sample. Among them, there are 106 sources that were falsely identified as AGN with a photometric redshift of $z \geq 3.5$. The false-positive rate of the AGN class for the whole 4XMM sample was expected to be about 5\%. This value was derived from the performance of the classification algorithm on a well-characterized validation sample, as detailed in the XMM2ATHENA project\footnote{http://xmm-ssc.irap.omp.eu/xmm2athena/wp-content/uploads/2023/04/DeliverableD8.2.pdf}. However, in the high-redshift regime, this increased to 10.8\%, highlighting the limitations of the algorithm in the case of rare sources such as high-z AGNs.

\subsection{Visual inspection of the images}\label{visual}

Section~\ref{data} describes the steps we took to obtain a clean sample by removing sources with unreliable photometry. Despite applying strict criteria, such as excluding objects near bright sources affected by diffraction spikes, some problematic photometry persisted in the high-z sample, leading to an artificially enhanced number of sources, particularly at $z > 6$. To address this, we visually inspected the images of all high-z sources across all bands (DES, VHS, WISE). This process enabled us to remove 32 sources that were strongly affected by nearby very bright stars or satellite trails, as their photometric redshift estimates were biased in one or more bands. The majority of these sources had falsely estimated photometric redshifts above $z \geq 6$.

%%%%%%%%%%%%%%%%%%%%%%%%%%%%%%%%%%%%%%%%%%%%%%%%%%%%%%%%%%%%%%%%%%%%%%%%%
%%%%%%%%%%%%%%%%%%%%%%%%%%%%%%%%%%%%%%%%%%%%%%%%%%%%%%%%%%%%%%%%%%%%%%%%%

\section{Final sample of the high-z X-ray sources}\label{final}

\begin{table}
\caption{Final sample of X-ray sources inside the 4XMM-DES-VHS footprint in different redshift bins after excluding stellar objects (Sect.~\ref{stars}) and sources with unreliable photometry (Sect.~\ref{visual}).}              % title of Table
\label{numbercounts}      % is used to refer this table in the text
\centering                                      % used for centering table
\begin{tabular}{c | r   r }          % centered columns (4 columns)
\hline\hline                        % inserts double horizontal lines
Redshift  & Number with    & Total \\
bins   & spec-z & number  \\
\hline                                % inserts single horizontal line
   Full sample &  13,424 (20.2\%)    &  64,529 \\
   $z\geqslant3.5$ & 92 (11\%)   &  833  \\
    $z\geqslant4$  & 30 (9.3\%)  & 323  \\
    $z\geqslant5$  & 3 (5.4\%)    & 56 \\ 
    $z\geqslant6$  & 0    & 7 \\ 
\hline                                             %inserts single line
\end{tabular}
\end{table}

\begin{figure}
\begin{center}
\includegraphics[width=0.47\textwidth]{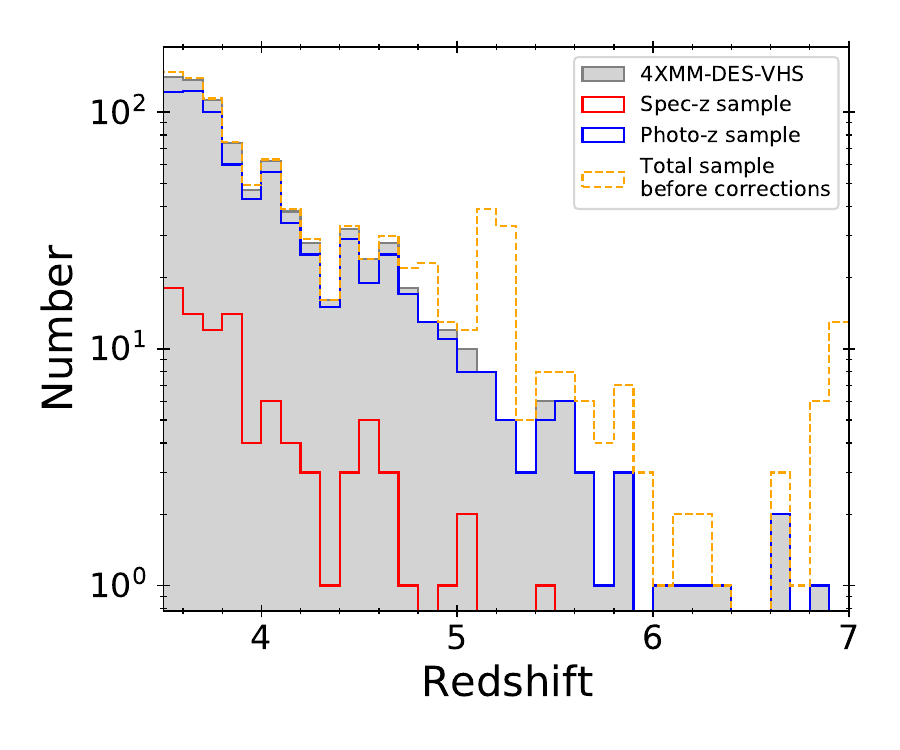} 
\end{center}
\caption{The redshift distribution of the final high-z sample (gray filled). We highlight the spectroscopic and photometric redshift samples as indicated in the legend. The dotted orange histogram shows the initial selection of high-z sources before removing stellar objects and sources with unreliable photometry.}\label{hist_highz} 
\end{figure}

Through our analysis, we selected in total 833 reliable X-ray sources in the early Universe ($z \geq 3.5$). Among them, there are 92 sources ($\sim$11\%) with available spectroscopic information. In Table~\ref{numbercounts}, we summarise the number of 4XMM-DES-VHS sources in different redshift intervals. Figure~\ref{hist_highz} shows the redshift distribution of the final high-z X-ray sample. In the same plot, we also display the samples separately, distinguishing those with photometric and spectroscopic redshifts. For comparison, we also present the distribution of the sample presented in Sect.~\ref{identificationhighz}, before the corrections in sections \ref{lowz}-\ref{visual}. The complete table of high-z X-ray sources is available at the CDS.

\subsection{Cumulative number counts}\label{sec:lognlogs}

\begin{figure}
\includegraphics[width=0.47\textwidth]{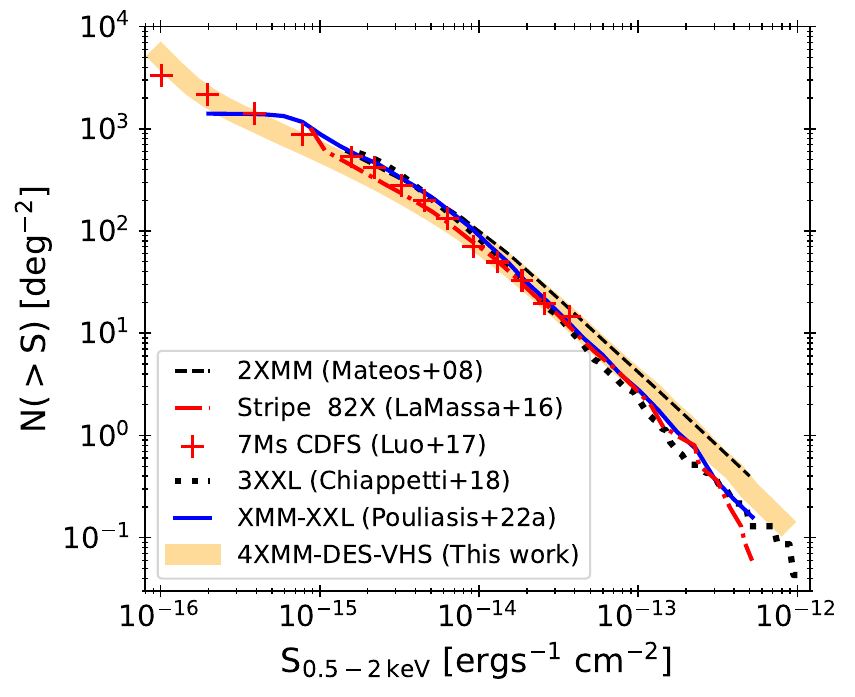}
\caption{The cumulative number counts for the 4XMM-DES-VHS catalogue are presented with the orange shaded area highlighting the 1-$\rm \sigma$ error in the soft 0.5-2 keV band. For reference, we compared with the number counts derived by \citet{Mateos2008}, \citet{lamassa2016}, \citet{luo2017}, \citet{Chiappetti2018} and \citet{Pouliasis2022a}.}
\label{integral}
\end{figure}

\begin{figure}
    \includegraphics[width=0.48\textwidth]{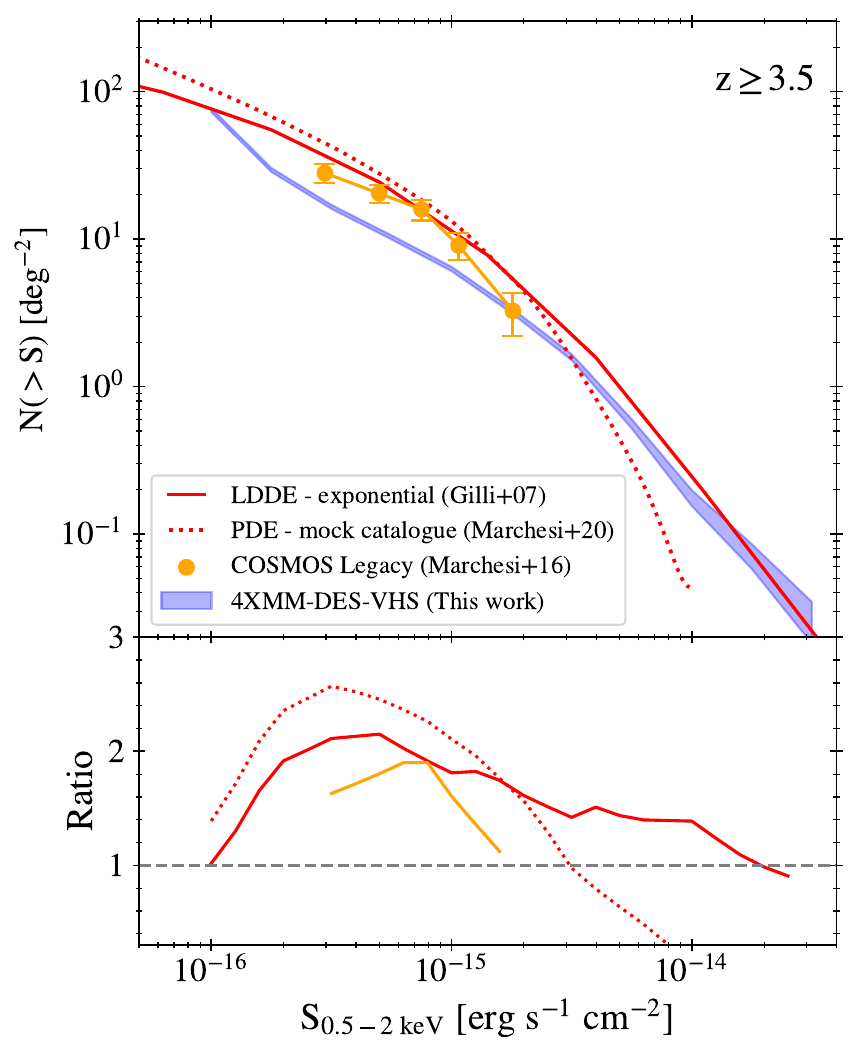}\\
\caption{The cumulative number counts for the high-z sample in the redshift bin $z\geqslant3.5$ (upper panel). The shaded blue area represent the 1$\rm \sigma$ uncertainties in our measurements. The solid line indicates the LDDE model predictions with an exponential decline by \citet{gilli2007}, while the dotted line shows the mock catalogue of \citet{Marchesi2020} based on the XLF by \citet{Vito2014}. For reference, we include the data points derived by \citet{marchesi2016}. The lower panel presents the ratio between our data and the different models.}\label{logNlogS}
\end{figure}

To verify that the number of high-z sources selected in our analysis aligns with the ones predicted by different models \citep[e.g.][]{gilli2007}, we calculate the cumulative number counts (logN-logS relation) and compare directly with the predictions. In order to calculate the logN-logS we need an estimation of the area curve of the 4XMM within the DES-VHS footprint. The XMM Survey Science Center provides EPIC sensitivity maps in the 0.2-12 keV band for each observation included in the 4XMM catalogue (see Sect.~9 of \citealt{Webb2020} for details about the calculation).\footnote{\url{http://xmm-catalog-dev.irap.omp.eu/~mcoriat/shared2/epfluxmap/}} A sensitivity map can be converted into an area curve by calculating the cumulative number of pixels in the map with a sensitivity limit above a certain X-ray flux. Since theoretical estimates of the logN-logS and previous observational results are usually presented in the 0.5-2 keV range, we converted the sensitivity maps into the same energy band by assuming the spectral model used for the calculation of the maps: an absorbed power law with photon index of $\Gamma = 1.42$ and a the column density of $\rm N_H=1.7\times10^{20}~\mathrm{cm}^{-2}$. The choice of the $\rm \Gamma$ and $\rm N_H$ follows the approach used in \citet{Webb2020} for constructing EPIC sensitivity maps. This spectral model is motivated by previous studies \citep[e.g.][]{Lumb2002} that characterized the unresolved extragalactic X-ray background, which is primarily composed of faint AGN. The photon index of 1.42 is a standard value representing the average spectrum of these sources, making it a reasonable choice for estimating sensitivity across the survey area. The chosen $\rm N_H$ value corresponds to a typical low level of Galactic absorption, ensuring that the sensitivity calculations are not significantly biased by intervening gas and dust. This standard approach allows for a consistent comparison with previous observational and theoretical logN-logS distributions. The overall footprint of the 4XMM in the DES-VHS area can be quite complex, with multiple overlapping XMM-Newton observations. However, by using the 4XMM detections catalogue for our logN-logS calculation, we do not need to consider this issue. In such case, each observation can be treated as an independent survey and the total area curve can be estimated as the sum of the area curve for each observation. Our final area curve for the 4XMM-DES-VHS includes 1,985 XMM-Newton observations.

As a sanity check of our method, we calculated the total logN-logS of the survey in the 0.5-2 keV band using our estimated area curve and the 106,790 4XMM-DR11 detections included in that sky region. The 0.5-2 keV flux for each detection was calculated by adding the band 2 (0.5-1 keV) and band 3 (1-2 keV) EPIC fluxes (\texttt{EP\_2\_FLUX} and \texttt{EP\_3\_FLUX} columns from the 4XMM catalogue). Figure~\ref{integral} shows the cumulative numbers derived using the described method (orange shaded region) compared to previous studies. In particular, we over-plot the logN-logS trends derived in the 2XMM \citep{Mateos2008}, the 3XMM-XXL and 4XMM-XXL surveys \citep{Chiappetti2018,Pouliasis2022a}, the 7 Ms Chandra deep field south \citep[CDFS,][]{luo2017} and the Stripe 82X field \citep{lamassa2016}. The number counts agrees very well with the aforementioned studies indicating that the area curve obtained in our analysis is accurate.

By using the logN-logS predicted by the X-ray background synthesis model of \citet{gilli2007},\footnote{\url{http://www.bo.astro.it/~gilli/counts.html}} and our area curve for the 4XMM-DES-VHS survey, we estimated a total of 1,681 X-ray detections in the 3.5 to 7 redshift range for this survey. The 833 unique 4XMM sources in our high-redshift sample correspond to 1,066 detections. The number of high redshift sources we identified is roughly of the same order a magnitude, but significantly lower.

Figure~\ref{logNlogS} shows the cumulative source distribution of the high-z sample. We compare our results to the predictions of the X-ray background synthesis model by \citet{gilli2007}, which is based on the optical luminosity function parametrised with a luminosity-dependent density evolution (LDDE) model and an exponential decline at high-z (solid line). We also show the number counts from the mock catalogue of X-ray selected AGN generated by \citet[dotted line,][]{Marchesi2020}, based on the X-ray luminosity function (XLF) by \citet{Vito2014}, which assumes a pure density evolution \citep[PDE,][]{schmidt1968}. For reference, we include the number counts derived by \citet{marchesi2016} using a $\rm z \geq 3$ sample from the COSMOS Legacy survey. The lower panel of Fig.~\ref{logNlogS} shows the ratio between our data and the different studies, with the shaded blue area representing the (1$\rm \sigma$) uncertainties in our measurements.

The number counts derived for the 4XMM-DES-VHS catalogue are systematically lower than the predictions from the LDDE model by \citet{gilli2007} and the PDE-based mock catalogue of \citet{Marchesi2020} across the entire flux range. At fainter fluxes ($S_{0.5-2 \text{keV}} < 10^{-15}$ erg s$^{-1}$ cm$^{-2}$), our results show a significant deficit compared to these models by a factor of up to 2–3. At the bright end, our measurements are in relatively closer agreement but still fall slightly below both models. A similar discrepancy is observed when comparing to the results of \citet{marchesi2016}. These differences are likely due to the incompleteness of our high-z sample at faint fluxes, as discussed in the next section.

\subsection{Completeness}\label{completeness}

\begin{figure}
\begin{center}
\includegraphics[width=0.45\textwidth]{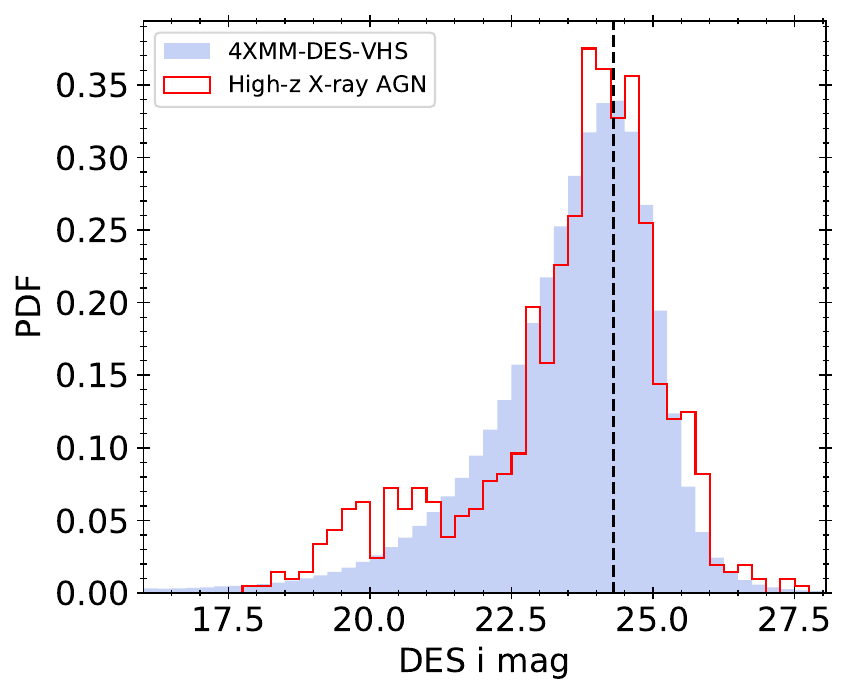} \\ 
\includegraphics[width=0.45\textwidth]{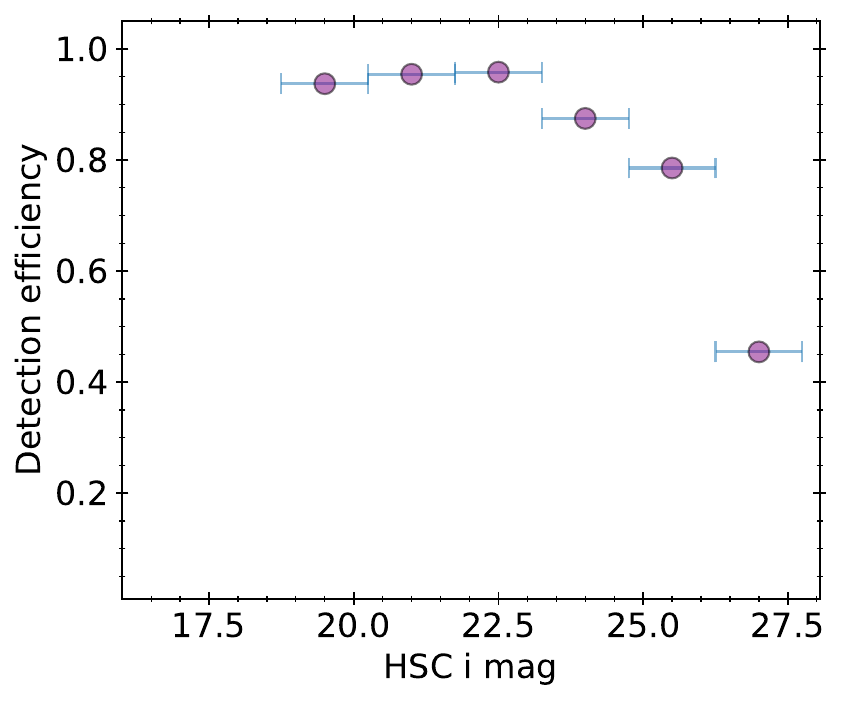} \\ 
\end{center}
\caption{Upper panel: Normalized histograms of the high-z X-ray AGN selected through our analysis (red) and the DES sources inside the 4XMM-DES-VHS footprint (blue) as a function of the DES i-band magnitude. The vertical line indicates the mode of the DES i-band magnitude distribution for the latter at $\rm i_{DES}=24.3~mag$. Lower panel: Fraction of high-z X-ray HSC sources that have a match with the DES DR2 catalogue.}\label{mag_i} 
\end{figure}

Figure~\ref{mag_i} (upper panel) shows the normalised histogram of the DES i-band magnitude for the high-z X-ray AGN compared to the full DES sample inside the 4XMM-DES-VHS footprint. The vertical line indicates the mode of the DES i-band magnitude distribution for the latter at $\rm i_{DES}=24.3~mag$. \citet{YangShen2023}, using the deeper ($\rm i_{HSC,limit}=26.3~mag$) Hyper-Suprime Camera PDR2 deep/ultradeep data \citep[HSC-PDR2,][]{Aihara2019}, derived the detection efficiency curves for all bands. The detection efficiency is defined as the fraction of the HSC objects that have a match in the DES DR2 catalogue (using a matching radius of 1"). In the i-band, the 95\% completeness detection limit is at $\rm i_{DES}=24.0~mag$, while at $\rm i_{DES}=25~mag$ this value drops to 40\%. Since our high-z sample peaks at about $\rm i_{DES}=24.3~mag$, we expect to miss sources at fainter magnitudes.

To quantify the detection efficiency of the high-z 4XMM-DES-VHS sources, we compared our sample to the high-z sample derived in the XMM-XXL-Northern field using the HSC data \citep{Pouliasis2024}. The latter study utilized the stacked XMM-Newton catalogue, which combines overlapping observations to achieve greater sensitivity and depth compared to the standard 4XMM-DR11 catalogue used in our analysis. Specifically, the stacked catalogue reaches a depth of $\rm 6 \times10^{-15}  \,erg~cm^{-2}~s^{-1}$ (at 3$\rm \sigma$) in the soft band ($\rm 0.5-2 keV)$, whereas the standard catalogue has a sensitivity limit of $\rm \sim4 \times10^{-14}  \,erg~cm^{-2}~s^{-1}$ at the same significance level. This difference in depth allows the stacked catalogue to probe fainter X-ray sources, particularly in the high-redshift regime.

Following a similar procedure as in \citet{YangShen2023}, but only considering sources with $\rm z \geq 3.5$, we matched the DES DR2 sources to the HSC data and restricted our sample to sources with X-ray fluxes above the limit of the 4XMM-DR11 catalogue. We then calculated the detection efficiency of the high-z sources in several magnitude bins, as shown in Fig.~\ref{mag_i} (lower panel). As expected, the completeness remains relatively constant at 90-95\% up to $i_{DES}=25~mag$ and then drops sharply. The total completeness with respect to the HSC high-z sample is about 86\%. Due to this incompleteness arising from the DES magnitude limit, we miss sources with faint X-ray fluxes, as seen in Fig.~\ref{x_compl}. In particular, at fluxes fainter than $\rm f_{0.5-2~keV} < 10^{-15} erg~s^{-1}~cm^{-2}$, we miss about 20-40\% of the sources. This explains the underestimation of the logN-logS relation at faint fluxes (Fig.~\ref{logNlogS}). The estimate of missed sources is derived from the decline in the fraction of 4XMM-DES-VHS sources with optical/near-IR counterparts (Fig.~\ref{x_compl}).

%%%%%%%%%%%%%%%%%%%%%%%%%%%%%%%%%%%%%%%%%%%%%%%%%%%%%%%%%%%%%%%%%%%%%%%%%
%%%%%%%%%%%%%%%%%%%%%%%%%%%%%%%%%%%%%%%%%%%%%%%%%%%%%%%%%%%%%%%%%%%%%%%%%

\section{Comparison with the optically-selected QSOs}
\label{sec:opticalqso}

\begin{figure}
\begin{center}
\begin{tabular}{c} 

\includegraphics[width=0.4\textwidth]{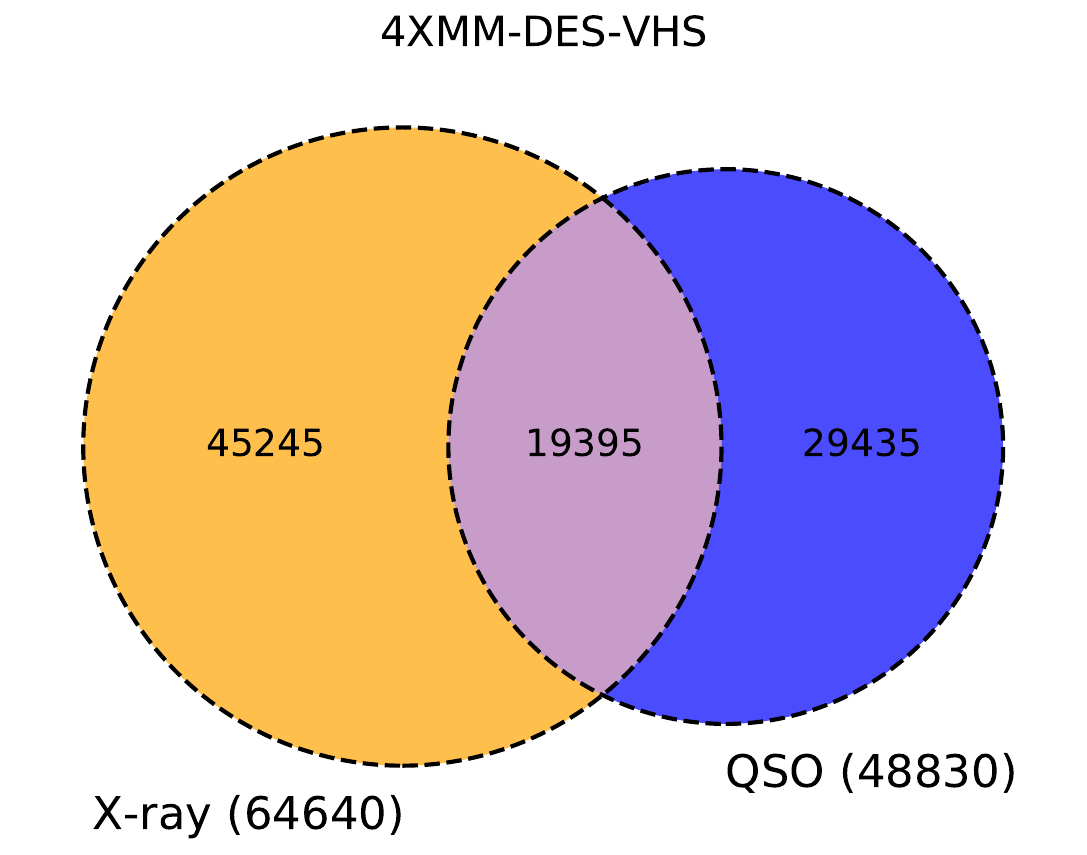}  \\
\includegraphics[width=0.4\textwidth]{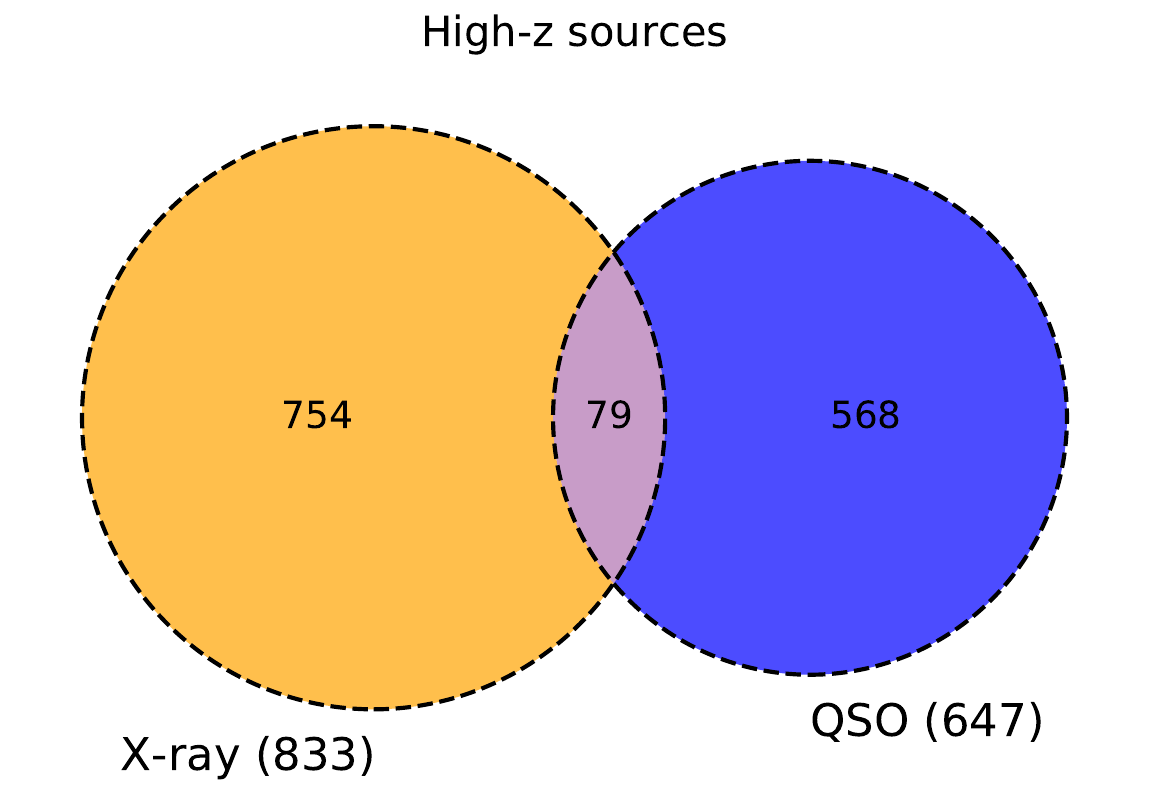}
\end{tabular}
\end{center}
\caption{Venn diagram of the X-ray sample used in our analysis and the optically-selected QSO candidates for the total (upper panel) and the high-z (lower panel) samples inside the 4XMM-DES-VHS area.}\label{venn} 
\end{figure}

One of the main goals of this work is to study the observational properties of the X-ray sources compared to the optically-selected QSOs in the early Universe. For the latter, we used a catalogue of 1.4 million photometrically-selected QSO candidates \citep{YangShen2023} in the southern hemisphere over the 5,000 $\rm deg^2$ DES DR2 survey. The QSO candidates are expected to be broad-line (unobscured) AGN with 94.7.\% reliability. Inside the 4XMM-DES-VHS footprint, there are 48,830 photometrically-selected QSOs. Among them, there are 686 sources with redshift $z \geq3.5$. As can be seen from the Venn diagram (Fig.~\ref{venn}), when comparing all the sources inside the area of our interest, about 30\% of the X-ray AGN are also broad-line AGN candidates, while $\sim$40\% of the QSOs have been detected in X-rays. Concerning the high-z samples, there is only a small portion of X-ray sources that are selected in the QSO sample and vice versa. In particular, only 9.5\% of the high-z X-ray AGN are also optically-selected QSOs. On the other hand, 12.5\% of the QSOs are X-ray detected. In the next subsections, we compare the X-ray properties, the magnitude distributions and the rest-frame SED shapes of both X-ray and optically-selected QSOs.

\subsection{X-ray properties}\label{sec:xrays}

\begin{figure}
\begin{center}
\includegraphics[width=\linewidth]{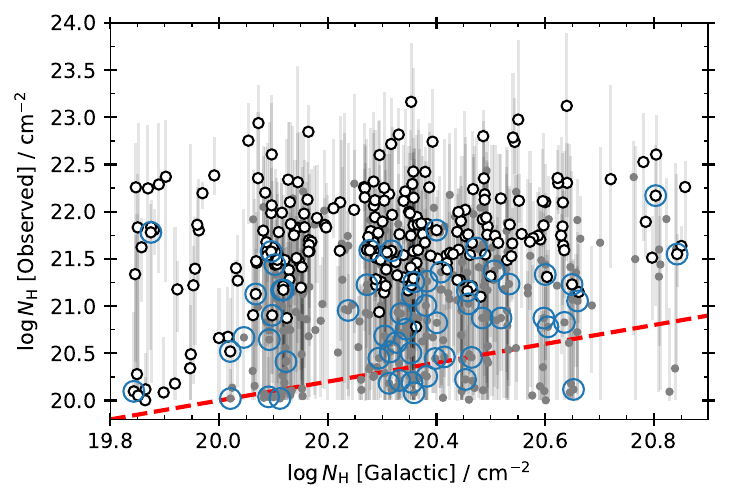} 
\end{center}
\caption{Galactic \nh\ versus the observed \nh\ for our high-z sample, estimated from the XMM2ATHENA 4XMM catalogs of X-ray spectral properties. The red, dashed line shows the one-to-one relation. Grey, solid circles are sources with an observed \nh\ consistent with the galactic {\nh}. The black, open circles show sources with \nh\ robustly above the Galactic value (the lower limit of the 90\% confidence interval is higher than the Galactic {\nh}). The blue circles mark the optically selected QSO identified by \citet{YangShen2023} also included in our sample. The plot only includes sources where our estimated 90\% confidence interval for the observed \lognh\ is narrower than 3 dex (501 out of 801 sources).}
\label{galvsobslognh} 
\end{figure}

\begin{figure}
\begin{center}
\includegraphics[width=\linewidth]{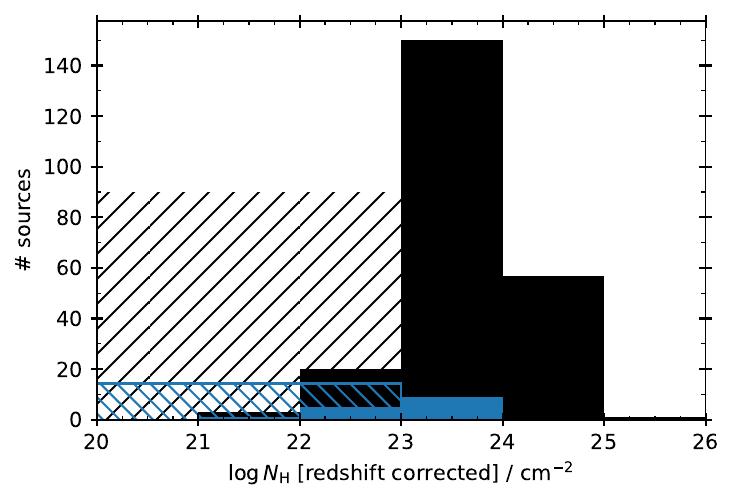} 
\end{center}
\caption{Estimated distribution of the rest-frame Hydrogen column density for our high-z sample. The black, solid histogram corresponds to sources with a significant observed \nh\ (i.e. above the Galactic value), and corrected to rest-frame. The hatched histogram shows our estimated number of unabsorbed sources. The blue histogram shows the corresponding distributions for the optically selected QSO by \citet{YangShen2023} included in our sample.}
\label{lognh_dist} 
\end{figure}

In this section, we present a brief view of the X-ray properties of our final high-z sample, and in particular, focusing on their X-ray absorption properties. To this end, we have used the catalogues of X-ray spectral properties built by the XMM2ATHENA project for the 4XMM sources.\footnote{\url{http://xmm-ssc.irap.omp.eu/xmm2athena/catalogues/}} These catalogues fit a simple X-ray spectral model (an absorbed power-law) to the 4XMM detections and derived estimates and uncertainties for the line-of-sight hydrogen column density, photon index, and 0.2-12 keV flux. The fits were done without any knowledge of redshift, so in the case of redshift dependent properties (e.g. \lognh) the quoted values in the catalogue should be considered as "observed" quantities and should be corrected to rest-frame if possible. For a detailed description of these catalogues see Viitanen et al. (in prep.).

We first searched for counterparts of our sample in the D6.1 XMM2ATHENA catalogue, which contains X-ray spectral properties for the 319,565 detections with at least one X-ray spectrum in the XMM-Newton archive. We found information for 161 detections, corresponding to 125 sources in our sample. After rejecting 27 detections flagged as bad fits in the catalogue (most are rejected due to bad background fits), we ended up with 104 sources (134 detections) with spectral properties from D6.1. These detections have a median net count of 90 and a signal-to-noise ratio of approximately 9. For the remaining objects in our sample we used the D6.3 XMM2ATHENA catalogue, which includes information for all the detections in the 4XMM. In this case, the X-ray spectral properties were derived by constructing low-resolution spectra using the 4XMM count-rates in the five standard XMM-Newton energy bands. We found 840 detections for the remaining 729 sources in our sample. We rejected 37 detections flagged either as potentially spurious or as bad fits, ending with a total of 698 sources (805 detections) with data from D6.3. This subset has a median net count of 30 and a signal-to-noise ratio of 6.

We calculated the average \lognh\ value and confidence interval for those sources with multiple detections. We, finally, rejected sources where our estimated confidence interval was larger than 3 dex, since we considered that no meaningful information about the X-ray absorption could be inferred in such cases. In all, from our sample of 833 high-redshift sources, we were able to obtain reasonably robust estimates of X-ray spectral parameters for 501 sources.\footnote{If we apply a more conservative criterion and select sources with a confidence interval no larger than 2 dex, the size of the final sample is reduced to 354 sources. This however does not affect significantly our conclusions about the fraction of absorbed sources.}

Our goal is to have a crude estimate of the fraction of sources that show signs of X-ray absorption. Since the spectral models used for the spectral fits in the XMM2ATHENA catalogues do not include the Galactic absorption, we must take this into consideration. Figure~\ref{galvsobslognh} shows our estimated observed \lognh\ average for the 501 high-z sources against the Galactic \nh\ towards their line-of-sight using the LAB Survey results \citep{Kalberla2005}. The grey circles are sources where the lower end of the \lognh\ confidence interval is below the Galactic absorption. We consider those sources (270 objects) as unabsorbed, since the observed \nh\ is consistent with the Galactic. On the other hand, the open white circles in the plot show sources with {\lognh} significantly above the Galactic absorption (231 objects). For these sources we can roughly estimate the rest-frame \nh\ taking into account the redshift information using the following expression:
\begin{equation}
\label{eq:nhzcorr}
\mnh(z) = \mnh(z=0) \times (1+z)^{2.7}.
\end{equation}

The black, solid histogram in Fig.~\ref{lognh_dist} shows the redshift-corrected \lognh\ distribution for these sources. Given the sensitivity range of XMM-Newton and the redshift of our sources, it is expected that their rest-frame \nh\ are above $\rm 10^{23}~\mathrm{cm}^{-2}$, since only sources with absorption roughly above that level would show significant absorption in their XMM-Newton spectra. We considered as X-ray absorbed those sources with \lognh$\ge23$. The black hatched histogram shows the number of sources which we considered as unabsorbed since they do not show an observed \nh\ above the Galactic level. We just show them uniformly distributed between 20 and 23.

\begin{figure*}
\begin{center}
\begin{tabular}{ccc} 

\includegraphics[width=0.3\textwidth]{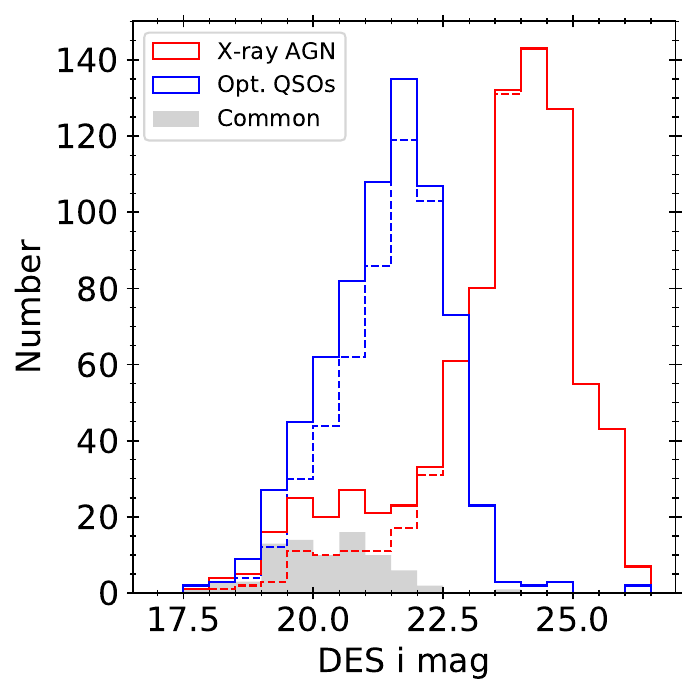}  &
\includegraphics[width=0.3\textwidth]{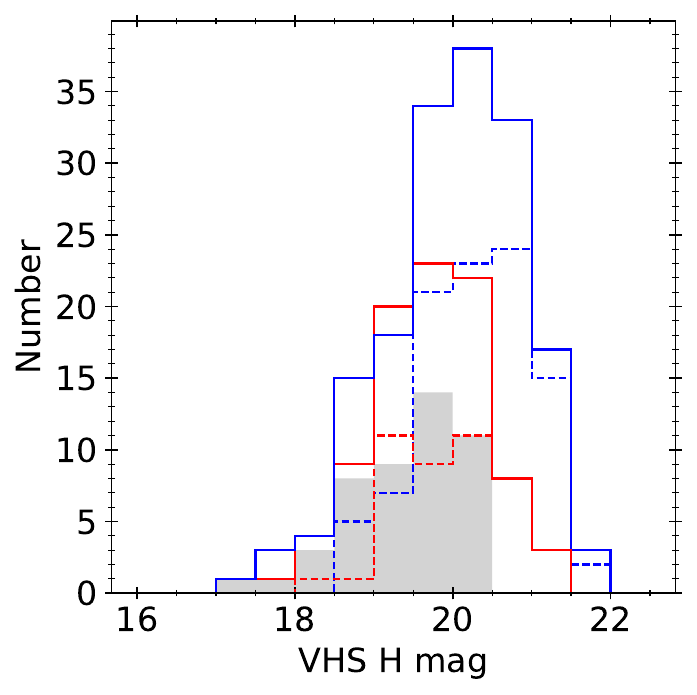} &
\includegraphics[width=0.3\textwidth]{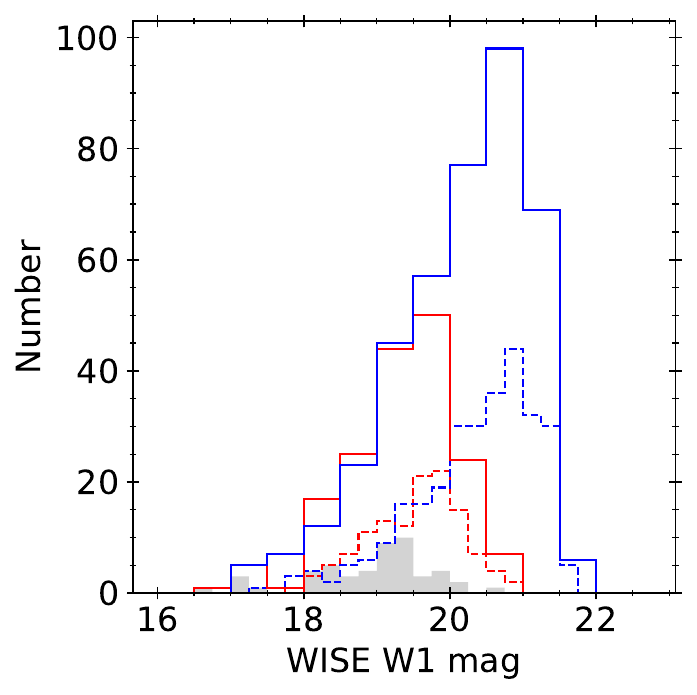} 
\end{tabular} 

\end{center}
\caption{The magnitude distributions of the X-ray AGN selected through our analysis (red) compared to the optically-selected QSO sample (blue) for the DES optical i-band (left), the VHS near-IR H band (middle) and the WISE mid-IR W1 band (right). The shaded histograms represent the common sources between the two samples, while the dashed lines correspond to the unique AGN in each selection method.}\label{mags} 
\end{figure*}
In total, out of the 501 sources with a reasonable \nh\ estimate, we classify 293 of them as X-ray unabsorbed and 208 as X-ray absorbed. The observed fraction of X-ray absorbed sources would be $40 \pm2 \%$. The uncertainties correspond to one-sigma confidence interval, calculated through bootstrapping. Note however that the uncertainties in the observed \lognh\ for a major number of our sources can be quite large (> 1 dex), which makes our estimate of the fraction of absorbed sources uncertain. When we applied a more conservative criteria for the identification of absorbed sources (e.g. the lower limit of the confidence limit for the redshift corrected \lognh\ is above 23), then we found 95 sources, corresponding to $20 \pm2 \%$ of X-ray absorbed sources. We rerun our pipeline using only the sources with spec-z information, and the derived fractions of obscured AGN remain the same. For a better identification of X-ray absorbed sources a detailed spectral analysis should be done, but that is beyond the scope of this work. Nevertheless, this crude estimate is in a reasonable agreement with previous studies. For example, \citet{Pouliasis2024} found an observed fraction of X-ray absorbed sources at high-z ($3 \leq z \leq 6$) in the XMM-XXL-Northern field (with an average depth comparable to the 4XMM) of $28.7$\%. 

It is also interesting to compare our results for the total sample with the optically-selected QSOs that are in common with our sample. In Fig.~\ref{galvsobslognh} the blue circles mark 58 sources identified as QSO in the DES survey \citep{YangShen2023}. The figure already suggests that the optical QSOs have observed \lognh\ towards the lower end of the distribution. By using the identification criteria for X-ray absorbed sources we presented above, we found nine absorbed sources ($\sim15$\%) out of the 58. Applying the more conservative criterion, only one QSO is classified as absorbed. As expected, the fraction of X-ray absorbed sources within optically selected QSOs seems to be significantly lower than in our total sample.

\subsection{Rest-frame Spectral Energy Distributions (SEDs)}\label{qso}

\begin{figure}
\begin{center}
\includegraphics[width=0.45\textwidth]{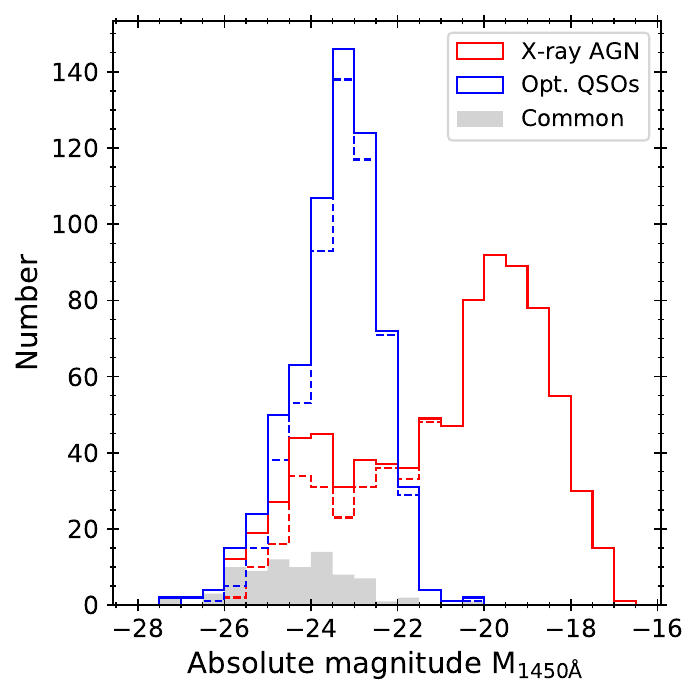}
\end{center}
\caption{The absolute magnitude distribution at $\rm 1450\AA$ of the X-ray AGN selected through our analysis (red) compared to the optically-selected QSO sample (blue). The shaded histograms represent the common sources between the two samples, while the dashed lines correspond to the unique AGN in each selection method.}\label{magsABS} 
\end{figure}

\begin{table*}
\caption{Models and their parameter space used by \texttt{CIGALE} for the SED fitting of the high-z sources.}
\begin{tabular}{ l c r }
\hline
\multicolumn{1}{l}{Parameter} &  & Value \\ \hline \hline
\multicolumn{3}{c}{Star formation history: delayed SFH with optional exponential burst)}\\
Age of the main stellar population in Myr && 50,100,200,500,1000,1500, 2000, 3000, 4000, 5000 \\
e-folding time of the main stellar population model in Myr, $\tau_{\rm main}$ && 50,100,200, 500, 700, 1000, 2000, 3000, 4000, 5000 \\
Age of the late burst in Myr, $age_{\rm burst}$ && 20  \\
Mass fraction of the late burst population, $f_{\rm burst}$ && 0.0, 0.005, 0.01, 0.015, 0.02,  \\
&& 0.05, 0.10, 0.15, 0.18, 0.20\\
e-folding time of the late starburst population model in Myr, $\tau_{\rm burst}$ && 50.0\\
\hline
\multicolumn{3}{c}{Stellar population synthesis model}\\
Single Stellar Population Library&&\citet{Bruzual2003}\\
Initial Mass Function&& \citet{Salpeter1955} \\
Metallicity && 0.02 (Solar) \\
\hline
\multicolumn{3}{c}{Nebular emission}\\
Ionization parameter ($\log U$)&& -2.0 \\
Fraction of Lyman continuum escaping the galaxy ($f_{\rm esc}$)&& 0.0 \\
Fraction of Lyman continuum absorbed by dust ($f_{\rm dust}$)&& 0.0 \\
Line width (FWHM) in km/s&& 300.0 \\
\hline
\multicolumn{3}{c}{Dust attenuation: modified attenuation law \citet{Charlot2000}} \\
V-band attenuation in the interstellar medium in mag, $Av_{ISM}$ && 0.2, 0.3, 0.4, 0.5, 0.6, 0.7, \\
&& 0.8, 0.9, 1, 1.5, 2, 2.5, 3, 3.5, 4\\
\hline
\multicolumn{3}{c}{Dust template: \citet{Dale2014}}\\
AGN fraction && 0.0\\
Alpha slope, $\alpha$ && 2.0\\
\hline
\multicolumn{3}{c}{AGN models from \citet{Stalevski2016} (SKIRTOR)}\\
 Average edge-on optical depth at 9.7 micron (t) &&   3.0, 7.0\\
 Power-law exponent that sets radial gradient of dust density (pl) && 1.0\\
 Index that sets dust density gradient with polar angle (q) && 1.0 \\
 Angle measured between the equatorial plane \\and edge of the torus in degrees (oa) && 40\\
 Ratio of outer to inner radius, $\rm R_{\rm out}/R_{\rm in}$ && 20\\
 Fraction of total dust mass inside clumps ($\rm M_{\rm cl}$) && 97\%\\
 Inclination angle in degrees ($i$) && 30, 70\\
 AGN fraction && 0.0, 0.1, 0.2, 0.3, 0.4, 0.5, 0.6, 0.7, 0.8, 0.9, 0.99 \\
 Extinction in polar direction in mag, E(B-V) && 0.0, 0.2, 0.4\\
 Emissivity of the polar dust && 1.6 \\
 Temperature of the polar dust (K) && 100.0\\
 The extinction law of polar dust && SMC\\

\hline
\end{tabular}
\tablefoot{Type-2 AGN have an inclination $i=70^\circ$, while type-1 AGN have $i=30^\circ$. The extinction in polar direction, E(B-V), included in the AGN module, accounts for the possible extinction in type-1 AGNs, due to polar dust. The AGN fraction is measured as the AGN emission relative to IR luminosity (1--1000 $\mu$m).}
\label{proposal}  
\end{table*}

\begin{figure}
\begin{center}
\includegraphics[width=0.5\textwidth]{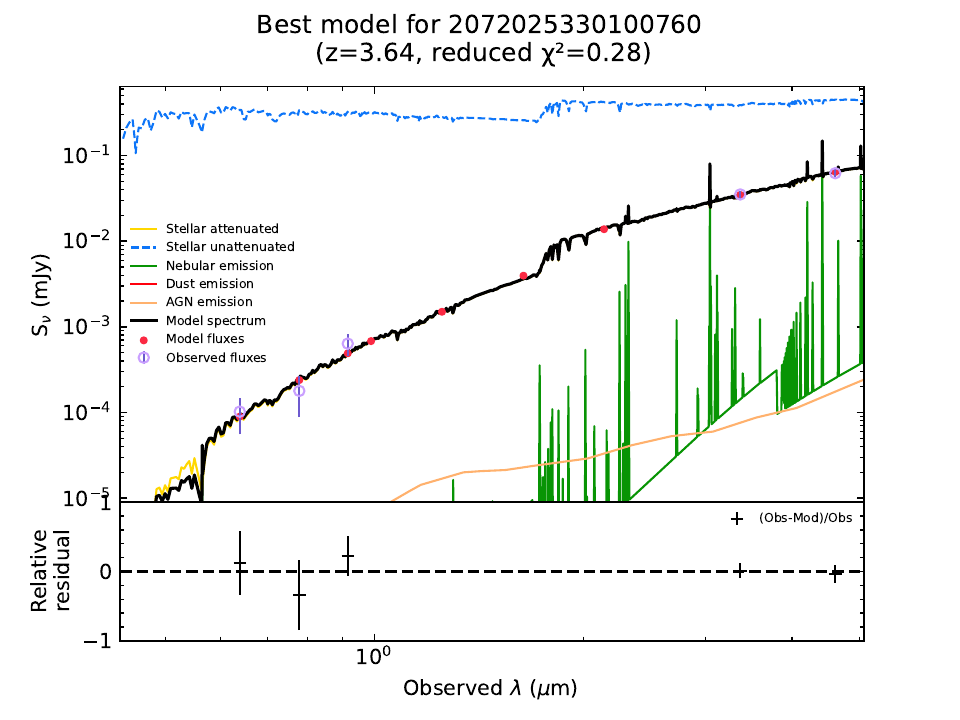}  \\
\includegraphics[width=0.5\textwidth]{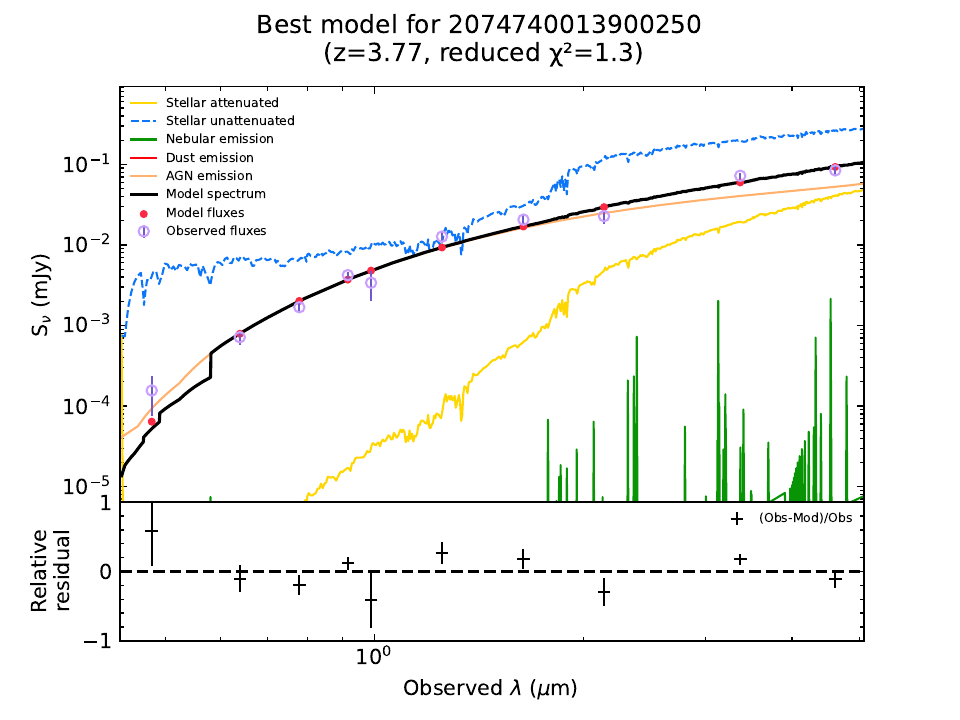}
\includegraphics[width=0.5\textwidth]{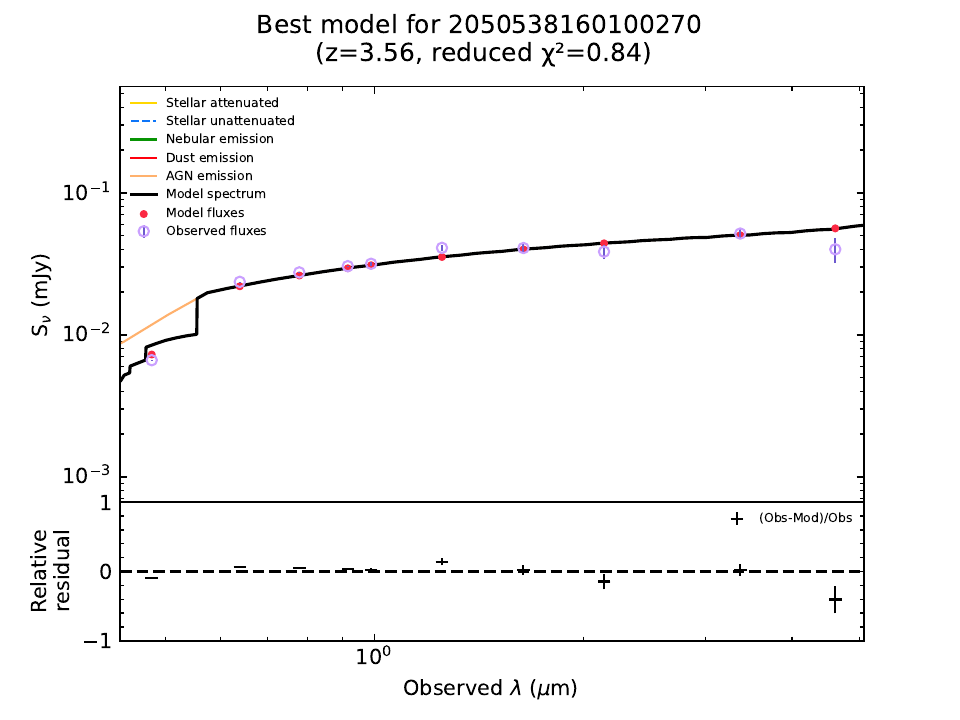}
\end{center}
\caption{Example SED fits of an X-ray absorbed AGN (top), an X-ray unabsorbed AGN (middle) and an optically selected QSO (bottom). The dust emission is plotted in red, the AGN component in orange, the attenuated (unattenuated) stellar component is shown with the yellow (blue) solid (dashed) line, while the green lines shows the nebular emission. The total flux is represented with black colour. Below each SED, we plot the relative residual fluxes versus the wavelength.}\label{exampleSED} 
\end{figure}

Figure~\ref{mags} presents the X-ray AGN (red) and optically-selected QSO (blue) distributions of the DES i-band (left), VHS H-band (middle) and WISE W1-band (right) magnitudes. The common sources between the two samples are shown also with the gray shaded histogram, while we overplot the X-ray AGN that are not optically-selected QSOs and vice-versa (dashed lines). The X-ray population covers the faint end of the DES i-band distribution, while the QSOs populate the bright one. One should take into account in this comparison that the selection function of the QSOs is biased towards the bright end in the optical. From the optical to mid-IR wavelengths, this trend is reversed; the QSOs have fainter W1 band compared to the X-ray sample. We also examined the absolute magnitude distribution at 1450Å for the same samples (Fig.~\ref{magsABS}). The absolute magnitudes were calculated after rerunning the LePhare algorithm using the final redshifts of the high-z sources. The QSO sample spans a range of approximately -26 to -22, with a median value of $\rm M_{1450 \AA}=-23.3$, while the X-ray AGN cover a much wider magnitude range from -26 to -18. However, the majority of the X-ray sources are at lower luminosities, with a median absolute magnitude of $M_{1450 \AA}=-20.3$. This confirms the observed trend that X-ray AGN tend to be fainter in the optical compared to optically selected QSOs, reflecting the bias of optical selection towards the brightest sources.

To understand better the above patterns, we aim to compare the rest-frame averaged SEDs of both the QSO and the X-ray AGN samples. We ran the \texttt{CIGALE} code \citep{Yang2020,Yang2022} to model and fit the SEDs of both the X-ray sample and the optically selected QSOs. \texttt{CIGALE} is a multi-component SED fitting algorithm that fits the observational data of the sources to the theoretical models, and has been widely used in the literature \citep[e.g.][]{Pouliasis2020}. We used all the available photometry to construct the SEDs, from optical to mid-IR, and we created a grid that models both the galaxy and the AGN emission.

In brief, we used the stellar population synthesis model of \citet{Bruzual2003} assuming the initial mass function (IMF) by \citet{Salpeter1955} and a constant solar metallicity (Z\,=\,0.02). We used the dust emission templates by \citet{Dale2014} without AGN emission and the dust extinction law by \citet{Charlot2000}. For the AGN component, we adopted the AGN templates presented in \citet[SKIRTOR]{Stalevski2012,Stalevski2016}. In addition for the star formation history (SFH), we used a delayed SFH with the functional form $\rm SFR\propto t\times exp(-t/\tau)$ that includes a star formation burst no longer than $\rm \tau = 20\,Myr$. We did not include the X-ray module in the SED-fitting process, as it requires intrinsic absorption-corrected X-ray fluxes, which cannot be reliably determined for our sample. Furthermore, since we are comparing the averaged SEDs of X-ray-detected AGN with optically selected QSOs (most of which lack X-ray detections), excluding X-ray fluxes ensures a consistent and unbiased comparison between the two populations. The list of the models and their parameters used in our analysis are given in Table~\ref{proposal}. With this configuration, we were able to fit the observational data with more than 650 million models. In Fig.~\ref{exampleSED}, we show three SED examples representative of the objects of our samples used in this work (e.g. X-ray obscured and unobscured AGN and an optically-selected QSO). In order to have results as reliable as possible, we kept all the sources that have low reduced $\rm \chi^2$ ($\rm \chi^2_r$) that is indicative of the goodness of the SED fitting process. Therefore, we excluded sources that have $\rm \chi^2_r>5$. This value has been used in previous studies \citep[e.g.][]{Mountrichas2019,Pouliasis2022b,Georgantopoulos2023} and is based on visual inspection of the SEDs. Following this criterion, we have excluded 36 X-ray AGN and 137 optically-selected QSO candidates from further analysis. We examined whether the poorly fitted SEDs correlate with the number of available photometric points, magnitudes, or redshift but found no clear dependence.

In \texttt{CIGALE}, the obscuration is primarily constrained by the inclination angle ($i$), with typical values of 30 degrees for type 1 AGN and 70 degrees for type 2 AGN \citep[e.g.][]{Mountrichas2022, Koutoulidis2022}. Additionally, for type-1 AGN it is possible to also account for dust extinction in the polar direction. The optical depth at 9.7 microns $\tau_{9.7}$ is not strongly constrained by the SED, as indicated by the flat relationship between the true and estimated values \citep{Yang2020, Yang2022}. Therefore, the classification into type 1 and type 2 AGN is based on the inclination angle, rather than $\tau_{9.7}$, with the algorithm showing a robustness of 70-85\% compared to spectroscopic samples in the literature \citep[e.g.][]{Mountrichas2021}. According to this criterion, 90\% of the QSO sample are classified as type 1 according to the SED fitting verifying the QSO selection criteria of \citet{YangShen2023}.

\begin{figure*}
\begin{center}
\begin{tabular}{cc} 
\includegraphics[width=0.47\textwidth]{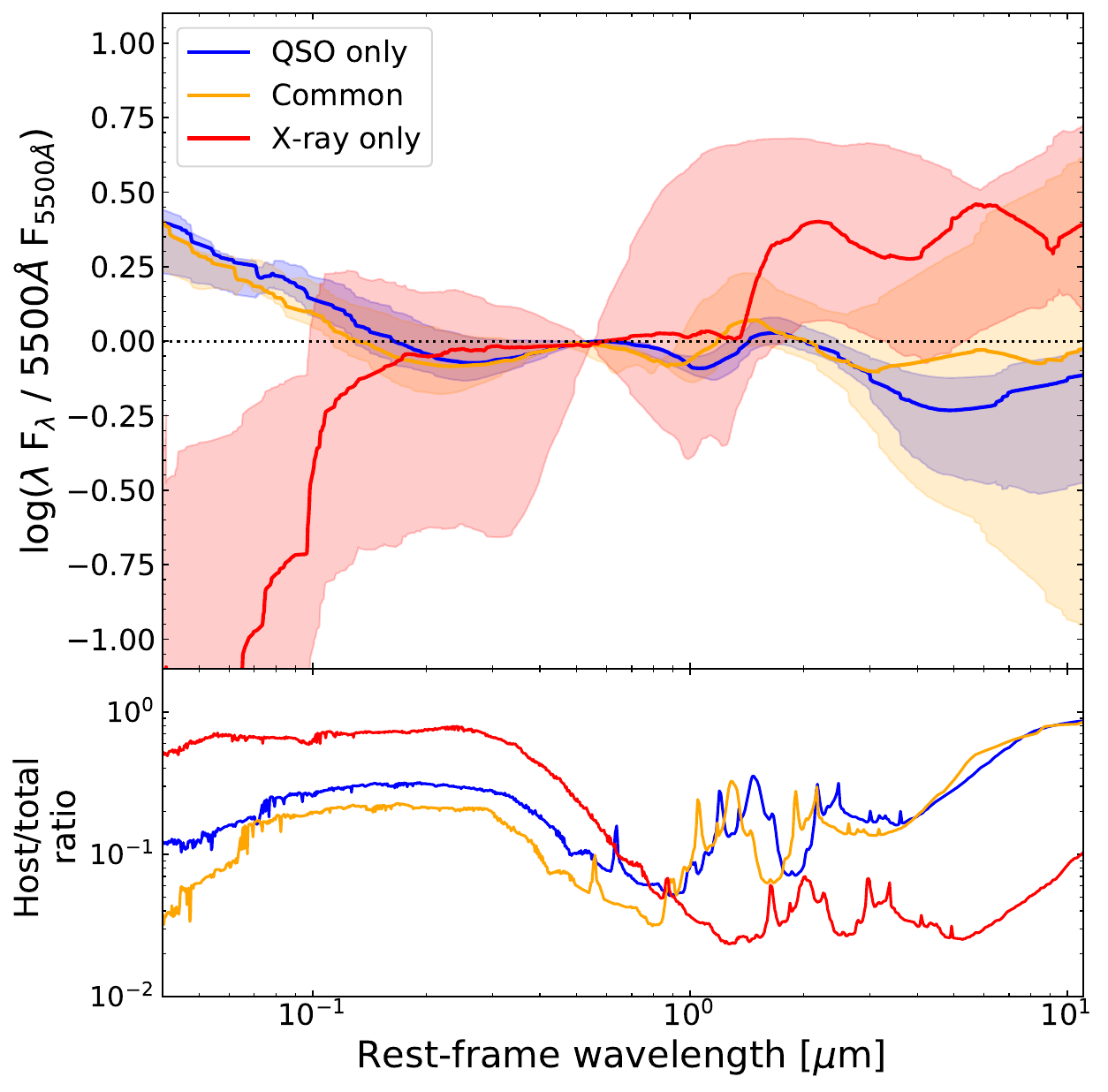} &
\includegraphics[width=0.47\textwidth]{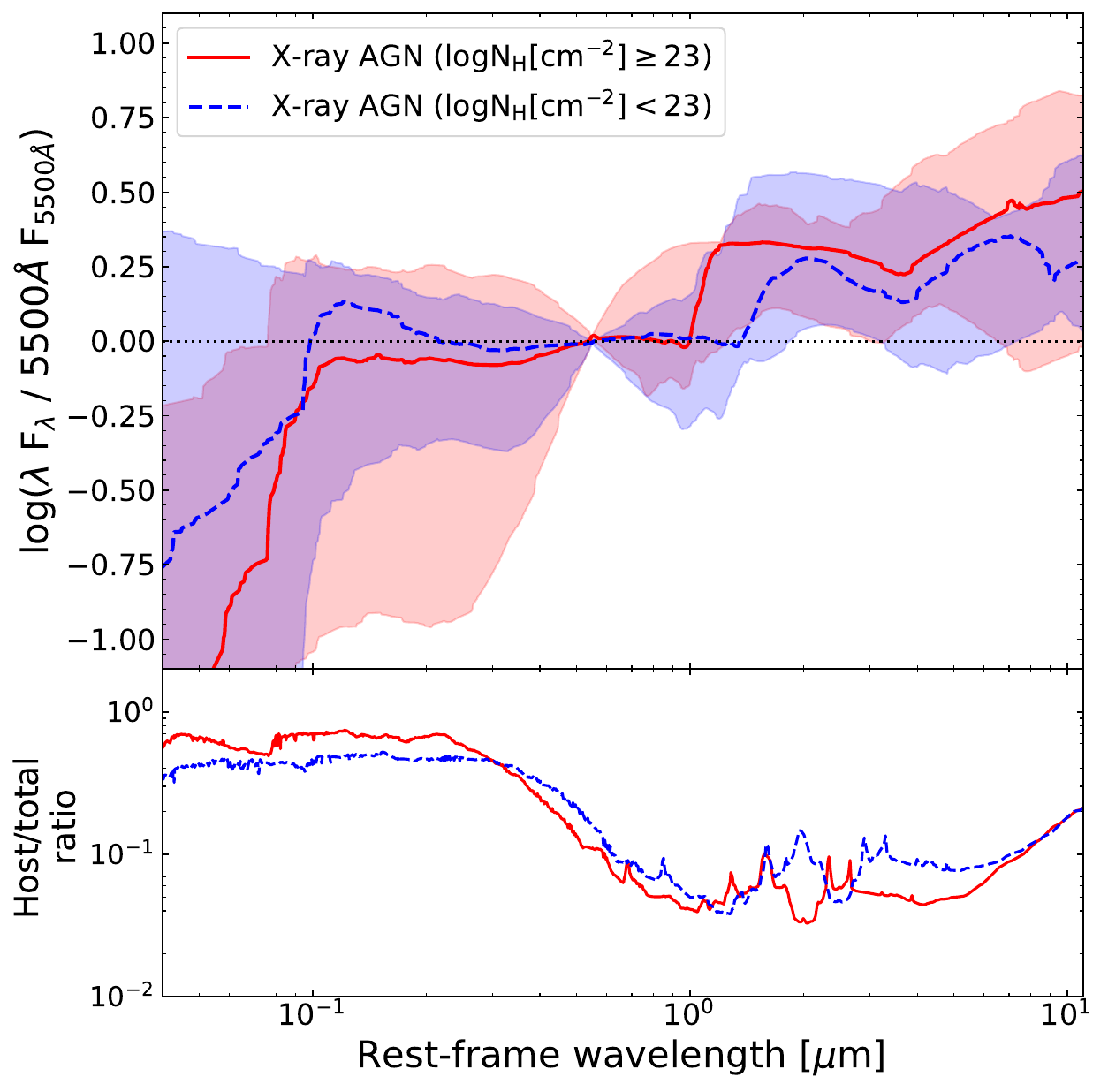} 
\end{tabular}
\end{center}
\caption{Left: Median SEDs (solid lines) of the X-ray (red) and optically (blue) selected AGN including the common sources (yellow). Right: Median SEDs (solid lines) of the X-ray selected AGN with $\rm \log N_H~(cm^{-2}) < 23$ (blue) and $\rm \log N_H~(cm^{-2}) \geq 23$ (red). The shaded regions correspond to the 25th and 75th percentiles of the dispersion of the templates at each wavelength. The bottom panel shows the ratio of the host-galaxy to the total emission. All individual SEDs are normalized at 5500$\AA$.}\label{sedX} 
\end{figure*}

The two plots in Fig.~\ref{sedX} provide a comparative analysis of the SEDs for different AGN populations, specifically optically-selected QSOs and X-ray-selected sources. These plots illustrate the efficacy of X-ray-selected samples in identifying AGN with varying levels of obscuration and showcase the distinct SED shapes of each population across the electromagnetic spectrum.

The upper left panel compares the median SEDs of QSOs (not X-ray detected, blue) and X-ray only-selected sources (red) across the rest-frame wavelength spectrum. The common sources in the two samples are presented in yellow colour. These SEDs are constructed by median-combining the individual SEDs of AGN in each sample, with the shaded regions marking the 25th to 75th percentile range to represent dispersion at each wavelength. For the QSO population, the SED in the UV-optical region is dominated by emission from the accretion disk, which is characteristic of unobscured AGN. This strong UV-optical emission, observed as a peak in the QSO SED, reflects minimal dust absorption and is a key indicator of the active central engine. In contrast, the X-ray-selected sources show a marked difference in SED shape. The UV-optical part of their spectrum appears significantly extincted, due to the presence of dust and gas obscuring the central engine. X-ray observations are particularly effective in detecting AGN across a broad range of obscuration levels, allowing for the identification of both heavily and lightly obscured sources, as evidenced by the absorption in this wavelength region for the X-ray population. In the infrared part of the spectrum, above 1 µm, the SED of the X-ray-selected AGN shows a noticeably higher level of emission compared to QSOs. This suggests that X-ray-selected AGN contain a larger amount of dust. The median SED for common sources sits between the QSO and X-ray AGN SEDs across much of the wavelength range. In the UV-optical part, common sources show evidence of accretion disk emission, similar to QSOs, indicating that they include unobscured AGN. In the infrared range,they show higher dust emission than QSOs, though still less than the purely X-ray-selected sources.

The lower left panel in Fig.~\ref{sedX} illustrates the ratio of host-galaxy to total emission across wavelengths, providing insights into the contributions from the AGN and host galaxy for each population. For the QSO-only sample, this ratio reveals as expected that the central engine overwhelmingly dominates the total emission, with minimal influence from the host galaxy. This aligns with the strong accretion disk emission observed in the UV-optical range, where the AGN outshines any host-galaxy contributions, resulting in a low host-galaxy to total emission ratio. In contrast, for X-ray only-selected sources, the emission is more balanced, with substantial contributions from both the AGN and the host galaxy. This balance reflects the impact of obscuration, which dims the central AGN and allows the host galaxy to be more visible. Consequently, the host-galaxy to total emission ratio is higher, yielding a flatter profile across wavelengths than seen in QSOs.

The right panel of Fig.~\ref{sedX} further explores these trends by comparing the median SEDs of high-z X-ray sources, specifically contrasting absorbed ($\rm \log N_H \geq 23$) and unobsorbed ($\rm \log N_H < 23$) AGN. Consistent with expectations, the median UV-optical emission of absorbed AGN is heavily obscured, whereas the unabsorbed AGN show less absorption and consequently slightly stronger emission in this range. The difference in X-ray obscured/unobscured AGN in the IR part of the spectrum does not seem so strong taking the median values. Also, when considering the 25-75 percentiles of the distributions (represented by the shaded regions in the plot), there is substantial overlap between absorbed and unabsorbed AGN. This overlap suggests that, despite the general trends, individual AGN can exhibit a wide range of SED shapes within each category, highlighting diversity within each population.

Hence, these plots demonstrate that X-ray-selected samples are highly effective in identifying AGN with varying levels of obscuration, capturing differences in SED shapes across both unobscured and obscured populations. The clear accretion disk emission in QSOs and common sources contrasts with the extinction observed in X-ray-selected sources in the UV-optical region, while the higher IR emission of X-ray AGN points to more substantial dust content.

\section{Summary and conclusions}\label{summary}

In this work, we made use of the 4XMM-DR11 catalogue (1240 $\rm deg^2$) that contains all the serendipitous X-ray sources from the \textit{XMM-Newton} space telescope in combination with the deep optical second release of the Dark Energy Survey (DES-DR2, 5000 $\rm deg^2$) to select the largest sample so far of high-redshift ($z \geq 3.5$) X-ray selected AGN. Our main goal was to study the properties of the high-z X-ray AGN and compare them to the high-z optically-selected QSOs. Our main results can be summarised as follows:

\begin{itemize}
    \item In the overlapping area of about 185 $\rm deg^2$ between the 4XMM-DR11, the DES-DR2 and VHSD-DR5 surveys there are 64,529 X-ray sources with a reliable optical or IR counterpart. For sources without spectroscopic information (about 80\%), we used the photometric redshifts derived using both SED fitting and machine-learning algorithms. After removing spurious sources, low-redshift galaxy interlopers and stellar objects contaminating our sample, we ended up with 833 high-z X-ray AGN with high reliability. In particular, the percentage of outliers between the spectroscopic and photometric redshifts is $\rm \eta \leq 12\%$ for $z \geq 3.5$. Among the high-z X-ray AGN, there are 92 sources (11\%) with available spectroscopic information.

    \item We were able to derive the logN-logS relation in the redshift bin $z\geqslant3.5$ at fluxes between $\rm f_{0.5-2~keV} = 10^{-16}-10^{-14}~erg~s^{-1}~cm^{-2}$. Our number counts are systematically lower than predictions from LDDE and PDE models, with a discrepancy of up to a factor of 2–3. These discrepancies are caused mainly by the incompleteness of our sample due to the sensitivity limits of the optical and near-IR bands.

    \item We compared the X-ray selected sample with the optically-selected broad-line QSO candidates in the same area. Only a small fraction ($\sim$ 10\%) of the high-z X-ray selected AGN are also optical QSOs and vice versa that is in contrast with the 35\% of the full sample. This result agrees with previous studies at high redshifts.

    \item By examining the X-ray properties of the sources, we found that among the X-ray sources the observed absorbed fraction ($\rm logN_H~[cm^{-2}] \geq 23$) is $\sim$20-40\%. This fraction is significantly lower (2-15\%) for the optical QSOs that are detected in X-rays (12.5\%).

    \item In comparing the magnitude distributions and rest-frame SEDs of QSOs and X-ray AGN, the X-ray AGN are observed to have fainter optical and brighter mid-IR magnitudes than optically-selected QSOs. Their rest-frame SED shapes also differ: optical QSOs are dominated by AGN emission in the UV/optical range with little host-galaxy influence, while X-ray AGN show significant absorption in these wavelengths for both absorbed and unabsorbed sources. Common sources serve as an intermediate group, showing moderate obscuration, with AGN emission more pronounced than in X-ray-selected sources.

\end{itemize}

We conclude that the differences among the AGN selection methods (optical, X-rays) are not due to selection or survey biases but on the differences of the intrinsically AGN or host-galaxy properties. Hence, it is crucial to use wide X-ray surveys (such as 4XMM) in order to uncover both the obscured (type 2) AGN population and the type 1 that are complementary to the optically-selected QSOs. Future, deeper and wider surveys, such as \textit{Euclid}, will increase the optical/IR identification of the X-ray sources and the completeness of the high-z AGN.

\section{Data availability}\label{availability}

The table with the final high-z X-ray-selected AGN is only available in electronic form at the CDS via anonymous ftp to cdsarc.u-strasbg.fr (130.79.128.5) or via http://cdsweb.u-strasbg.fr/cgi-bin/qcat?J/A+A/.

\begin{acknowledgements}
    The authors are grateful to the anonymous referee for a careful reading and helpful feedback. EP, AR, IG, AA and SM acknowledge financial support by the European Union's Horizon 2020 programme "XMM2ATHENA" under grant agreement No 101004168. The research leading to these results has received funding (EP and IG) from the European Union's Horizon 2020 Programme under the AHEAD2020 project (grant agreement n.871158). NW acknowledges support from the Centre national d'études spatiales (CNES) for this work. FC acknowledges funding from grant PID2021-122955OB-C41 funded by MCIN/AEI/10.13039/501100011033 and by ERDF A way of making Europe. This research made use of Astropy, a community-developed core Python package for Astronomy \citep[\url{http://www.astropy.org},][]{astropy2018}. This publication made use of TOPCAT \citep{Taylor2005} for table manipulations. The plots in this publication were produced using Matplotlib, a Python library for publication quality graphics \citep{Hunter2007}. Based on observations obtained with \textit{XMM-Newton}, an ESA science mission with instruments and contributions directly funded by ESA member states and NASA.
\end{acknowledgements}

\bibliographystyle{aa}
\bibliography{bibliography}

\end{document}